\documentclass[aps,twocolumn,showpacs,superscriptaddress,groupedaddress,nofootinbib]{revtex4} 
\usepackage{graphicx}
\usepackage[sort&compress]{natbib}
\usepackage{subfigure}
\usepackage{amsmath}
\usepackage{amssymb}
\usepackage{amsfonts}
\usepackage{rotating}
\usepackage{cancel}
\usepackage{lipsum}
\usepackage{mathtools}
\usepackage{color}
\usepackage{bbm}
\usepackage{dsfont}
\usepackage{bbold}
\usepackage{multirow}
\usepackage{ulem}

\newcommand{\eps}{\epsilon}

\begin{document}

\title{Quark mass dependence of the $D_{s0}^*(2317)$ and $D_{s1}(2460)$ resonances}
\author{F. Gil-Dom\'inguez}
\email{fernando.gil@ific.uv.es}
\affiliation{Departamento de F\'{\i}sica Te\'orica and IFIC,
Centro Mixto Universidad de Valencia-CSIC, Parc Científic UV, C/ Catedrático José Beltrán, 2, 46980 Paterna, Spain}  
\author{R. Molina}
\email{Raquel.Molina@ific.uv.es}
\affiliation{Departamento de F\'{\i}sica Te\'orica and IFIC,
Centro Mixto Universidad de Valencia-CSIC, Parc Científic UV, C/ Catedrático José Beltrán, 2, 46980 Paterna, Spain}

\begin{abstract}  
We determine the quark mass dependence - light and heavy - of the $D_{s0}^*(2317)$ and $D_{s1}(2460)$ properties, such as, mass, coupling to $D^{(*)}K$, scattering lengths and compositeness, from a global analysis of $D^{(*)}K$ energy levels from LQCD. In particular, we analyze the HSC energy levels for $DK$ scattering in $I=0$ for different boosts and two pion masses. The formalism is based in the local hidden-gauge interaction of Weinberg-Tomozawa type which respects both, chiral and heavy quark spin symmetries, supplemented by a term that takes into account the $D^{(*)}K$ coupling to a bare $c\bar{s}$ component. The isospin violating decay of the $D_{s0}^*(2317)\to D_{s}^+\pi^0$ is also evaluated.
\end{abstract}
\maketitle
\section{Introduction}
In order to describe the heavy meson spectrum, it is useful to think on the restrictions arising from the heavy quark limit ($m_Q\to\infty$). As it is well-known, the interaction becomes independent on the spin of the heavy quark in this limit due to the inverse color-magnetic-moment dependence on the heavy quark mass. Then, the heavy and light degrees of freedom are separately conserved and heavy mesons form mass doublets whose components are related by a flip of the heavy quark spin. Regarding the charmed-strange meson spectrum, if one assumes that the mass of the charm quark is sufficiently heavy to consider this limit, then, two degenerate states for $L=0$ emerge for the ground state with spin-parity $J^P=0^-,1^-$, being the total spin of the light degrees of freedom $J^P_l=\frac{1}{2}^-$. These states correspond naturally to the $D$ and $D^*$ mesons. On the other hand, for $l=1$, one has two doublets, the first one for $J^P_l=\frac{1}{2}^+$, $J^P=0^+,1^+$, which in principle might be related to the observed states $D_{s0}^*(2317)$ and $D_{s1}(2460)$ \cite{babards0,cleods1}, and the other for $J^P_l=\frac{3}{2}^+$, $J^P=1^+,2^+$,  could be associated to the $D_{s1}(2536)$ and $D_{s2}(2573)$. Still, the constituent quark model predicts the $J^P_l=\frac{1}{2}^+$ doublet to be almost degenerate in mass and quite broad decaying to $D^{(*)}K$ \cite{godfrey,godfrey2,dipierro}. Nevertheless, what is observed is a couple of narrow states whose masses differ in about the $m_{D^*}-m_{D}$ splitting, being the $D_{s0}^*(2317)$ and $D_{s1}(2460)$ very close to the $DK$ and $D^*K$ thresholds respectively. Moreover, these two states have the same binding energy relative to these thresholds, $\sim 45$ MeV. For these reasons, molecular explanations have been proposed \cite{barnesclose,kolomeitsevlutz,faesslergutsche,faesslergutsche2,lutzsoyeur,Gamermann:2006nm,Gamermann:2007fi}. From the experimental point of view, several new exotic states in the charm and strangeness sector have been observed in the last years. Remarkably, new states close to $D^*K^*$ threshold, $T_{cs}(2900)$ and $T_{c\bar{s}}(2900)$, have been recently observed, which were predicted earlier in \cite{molinabranz}. See also \cite{Molinatcs}. Consequently, the study of exotic states with charm and strangeness remains interesting.


From lattice QCD simulations, reasonable descriptions of both, the $D_{s0}^*(2317)$ and $D_{s1}(2460)$ have been obtained when $D^{(*)}K$ interpolators are included \cite{bali,mohler,lang,liuorginos,cheungthomas,alexandrouc}, while their masses are overestimated if omitted. In \cite{alberto}, the lattice energy levels of \cite{mohler,lang} are reanalyzed using an auxiliary potential for $D^{(*)}K$, which comes from a linear potential in the center-of-mass energy and a Castillejo-Dalitz-Dyson (CDD) pole \cite{Castillejo:1955ed}. This results in a couple of bound states with binding energy of $\sim 40$ MeV with respect to the $DK$ and $D^{*}K$ thresholds, with a probability of $70\%$ for the $DK$ and $D^{*}K$ components in the $D_{s0}^*(2317)$ and $D_{s1}(2460)$ states, respectively. This outcome encounters consistency with predictions for $I=0$ $DK$ scattering from previous analyses using Heavy Meson Chiral Perturbation Theory (HM$\chi$PT) for different sets of lattice data on scattering lengths for $I=3/2$ $D\pi$, $D_s\pi$, $D_sK$, $I=0,1$ $D\bar{K}$, with pion masses in the range of $300-550$ \cite{liuorginos}. In \cite{liuorginos}, the quark mass dependence of the scattering lengths are also extracted, and the isospin-breaking decay width $\Gamma(D_{s0}^*(2317)\to D_s\pi)$ is predicted to be $133\pm 22$ KeV being consistent within the current experimental upper limit of $3.8$ MeV. This decay seems to be very sensitive to the internal structure of the $D_{s0}^*(2317)$. If the $D_{s0}^*(2317)$ were a $c\bar{s}$ state, in ~\cite{Handecay}, using the Bethe-Salpeter equation to describe a relativistic bound state, a decay width of $\Gamma(D_{s0}^*(2317)\to D_s\pi=7.83^{+1.97}_{-1.55}$~KeV has been predicted (see also~\cite{Colangelo2003,Colangelo2005}). In contrast, the latest calculations in the molecular model, give $ 120^{+18}_{-4}$ KeV~\cite{Fudecay} for this decay (see also~\cite{Clevendecay}).

In \cite{menglin} the analysis of~\cite{liuorginos} is extended to leading one-loop order (next-to-next leading order or NNLO) in the charmed-meson light-meson scattering amplitudes including the $N_f=2+1$ lattice data of \cite{liuorginos} for pion masses in the range $300-620$ MeV and \cite{mohler} for $m_\pi=156$ MeV. However, with seven LECs to be fitted to the lattice data that were available ($\sim 20$ data points), the result of the LECs bear large uncertainties. In principle, the effect of the $c\bar{s}$ degree-of-freedom is encoded in the LECs beyond LO in the chiral expansion and in the resummation required by unitarity. Nevertheless, if these kind of configurations are close by the energy region under consideration, it could be more accurate to consider this degree of freedom explicitly \cite{pedro,Yang:2021tvc}. In \cite{pedro}, the lattice data of \cite{bali} are analyzed, and a reasonably good description can be obtained coupling the LO $D^{(*)}K$ interaction from HM$\chi$PT with $c\bar{s}$. The data set used supersedes \cite{mohler,lang} since \cite{bali} uses larger volumes in a finer lattice spacing and one order of magnitude more gauge configurations. However, in the $N_f=2$ simulation with dynamical fermions  of \cite{bali}, effects emerging from strange sea quarks are neglected. In \cite{pedro}, $DK$ and $D_s\eta$ channels are included in the analysis. Notably, similar results are obtained when NLO terms in the scattering amplitude are included, and also, taking into account two different renormalization schemes. The amount of the molecular component found is $70\pm 10\%$ (including both channels, $54-63\%$ corresponding to $DK$). This result is in contradiction to the one obtained by the constituent quark model (CQM) \cite{vijande,valcarce,segovia} which uses the $^3P_0$ transition operator \cite{leyaouanc} to couple the quark-antiquark and meson-meson d.o.f's. As a result, the dressed state had a $DK$ meson pair probability of around $33\%$. 

$DK$ femtoscopic correlation functions have been recently estimated~\cite{Liucor,Albaladejocor,Ikenocor}. These observables can also be useful to extract the interaction and scattering parameters, being sensitive to the nature of exotic hadrons~\cite{Ikenocor}. 

In~\cite{Clevenmass} a prediction of the pion and kaon mass dependence of the $D_{s0}^*(2317)$ is provided in the framework of NLO HM$\chi$PT. There, it is argued that, the light ($u,d$) quark mass dependence of the $D_{s0}^*(2317)$ mass should be different for a molecular state compared to a $c\bar{s}$ state. The main reason is that this dependence comes through meson-meson loops, which depend on the square of the coupling of the resonance to the pair of mesons, being this coupling much larger for molecular states. In addition, the loop function peaks right at the threshold, the molecular state following the threshold, which is linear in the meson masses. While the masses of the charmed mesons grow slightly with the squared of the pion mass~\cite{gil}, the mass of the kaon grows linearly with it, being the pion mass proportional to the square root of the light quark mass. However, for a compact state, the light quark mass dependence is always an analytic function in the light quark masses which is quadratic in the Goldstone-Boson masses. In \cite{Clevenmass} the type of pion mass dependence obtained for the $D_{s0}^*(2317)$ pole is quadratic, while being linear with slope close to $1$ for the kaon mass dependence.

Nonetheless, the overall quark mass dependence of the $D_{s0}^*(2317)$ and its properties have not been extracted from the available bulk of LQCD simulations yet. In this work we provide a LQCD bare analysis of the quark mass dependence of the $D_{s0}^*(2317)$ and $D_{s1}(2460)$. Concretely, the recent data of \cite{cheungthomas} that investigate $I=0$ $DK$ scattering for two pions masses, $m_\pi=239$, $391$ MeV are analyzed for the first time, comparing with previous lattice QCD analyses, and, finally, we perform a global fit of the available LQCD data \cite{lang,mohler, alberto,bali, cheungthomas}, extracting accurately the quark mass dependence, light and heavy, of the $D_{s0}^*(2317)$ parameters, and its degree of molecular state.
The meson-meson interaction is provided by the hidden gauge lagrangian, which at leading order coincides with the one of HM$\chi$PT supplied by a CDD pole term.

Compared with previous studies \cite{liuorginos,alberto,Clevendecay,menglin,pedro} this work considers the analysis of the recent data \cite{cheungthomas} and of all available LQCD data in $DK$, $D^*K$ scattering, extracting the overall quark mass dependence of the $D_{s0}^*(2317)$ and $D_{s1}(2460)$ properties.


\section{Formalism}
\begin{figure*}
   \begin{minipage}{0.23\textwidth}
     \centering
     \includegraphics[width=1.1\linewidth]{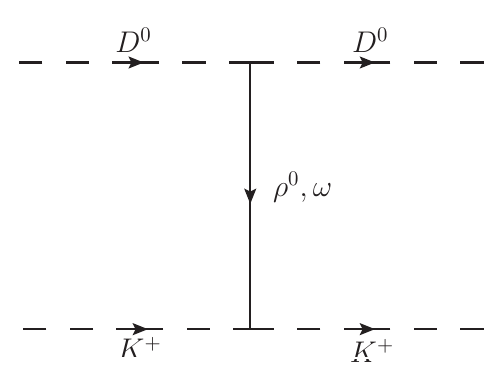}
   \end{minipage}
   \hspace{1cm}
   \begin{minipage}{0.23\textwidth}
     \centering
     \includegraphics[width=1.1\linewidth]{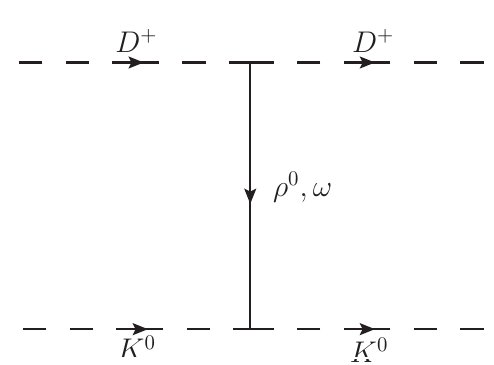}
   \end{minipage} 
   \hspace{1cm}
   \begin{minipage}{0.23\textwidth}
     \centering
     \includegraphics[width=1.1\linewidth]{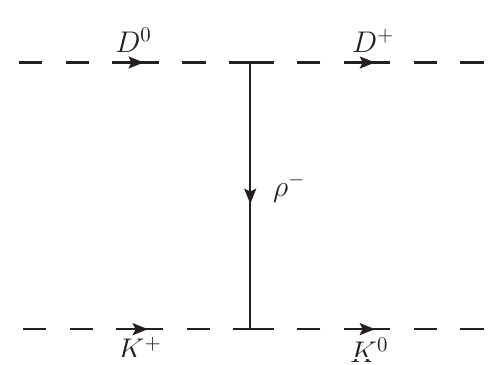}
   \end{minipage} 
   \caption{Feynman diagrams of the $DK\rightarrow DK$ process.}
   \label{fig:DKdiagrams}
\end{figure*} 
\begin{figure}
     \centering
     \includegraphics[width=0.8\linewidth]{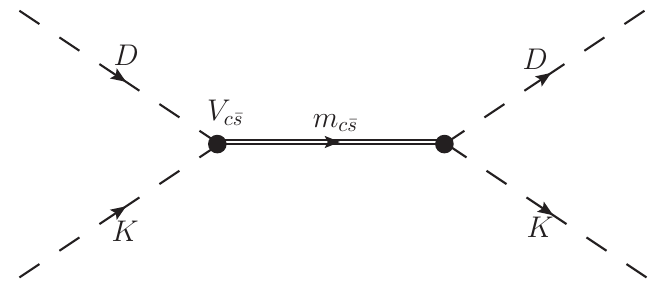}
   \caption{Feynman diagram of the $DK\rightarrow DK$ process with an interaction with a bare $c\bar{s}$ state.}
   \label{fig:DKmcsdiagram}
\end{figure} 
We consider that the $DK\to DK$ scattering is driven by vector meson exchange and evaluate it using the lagrangian from the Hidden Gauge Formalism (HGF) \cite{bando1,bando2,roca,Meissnereport},
\begin{eqnarray}
 {\cal L}=ig\mathrm{Tr}\left([\partial_\mu\phi,\phi]V^\mu\right)
 \label{eq:lag}
\end{eqnarray}
with $g=m_\rho/2f_\pi$, $f_\pi$ the pion decay constant and $m_\rho$ the $\rho$ meson mass. The matrices $\phi$ and $V$ contain the 16-plet of pseudoscalar and vector mesons and are given in the Appendix. This lagrangian is consistent with both, chiral and heavy quark spin symmetries, for the light vector-meson exchange, since there the heavy quark is acting as an spectator and the interaction is broken to SU(3) \cite{xiao}. Considering only light vector meson exchange, see Fig.~\ref{fig:DKdiagrams}, this interaction leads to,
\begin{eqnarray}
 V_{DK}=-\frac{s-u}{2f_\pi^2}, \label{eq:l1}
\end{eqnarray}
where $s$ and $u$ are the Mandelstam variables. Note that in the above equation the pion decay constant is not taken as a constant parameter, but its pion mass dependence is taken from \cite{Molina:2020qpw}. Note also that one arrives to this interaction through the LO HM$\chi$PT \cite{liuorginos,pedro}. A term associated with the interaction of $DK$ with a bare $c\bar{s}$ state of the $J^P_l=\frac{1}{2}^+$ Heavy Quark Spin Symmetry (HQSS) doublet can be added as shown in Fig.~\ref{fig:DKmcsdiagram}. At LO in the heavy quark expansion this gives~\cite{pedro},
\begin{eqnarray}
 V_\mathrm{ex}=\frac{V_{c\bar{s}}^2}{s-m_{c\bar{s}}^2}\ , \label{eq:l2}
\end{eqnarray}
with
\begin{eqnarray}
 V_{c\bar{s}}(s)=-\frac{c}{f_\pi}\sqrt{M_Dm_{c\bar{s}}}\frac{s+m_K^2-M_D^2}{\sqrt{s}}\ ,\label{eq:couy}
\end{eqnarray}
where $m_{c\bar{s}}$ is the mass of the bare $c\bar{s}$ component, $m_{c\bar{s}}=m_{c\bar{s}}^{0^+}$, and $c$ is a dimensionless constant that provides the strength of the coupling of this component to the $DK$ channel.
This constant was evaluated in~\cite{pedro} from a fit to the LQCD energy levels of the $b\bar{s}$ $J^P=\left\{0,1\right\}^+$, obtaining $c=0.74 \pm 0.05$. In this work we will consider this coupling as a free parameter, together with $m_{c\bar{s}}$.
\begin{table}[h!]
 \setlength{\tabcolsep}{0.5em}
{\renewcommand{\arraystretch}{1.6}
\centering
\begin{tabular}{|c|c|c|c|c|c|}
\hline
 Col. & $a$ & $L$ & $m_\pi$ & $m_{D_s}$ & $M_{avg}$ \\
 \hline
 \multirow{2}{*}{HSC} & \multirow{2}{*}{0.12} & 3.8 & 239 & 1967 & 3051 \\
& & $1.9-2.9$ & 391 & 1951 & 3024 \\
 \hline
 \multirow{2}{*}{RQCD} & \multirow{2}{*}{0.071} & $1.7-4.5$ & 150 & 1978 & 3068\\
     &  & $3.4-4.5$ & 290 & 1977 & 3068 \\
 \hline
 \multirow{2}{*}{Prelovsek et al.} & 0.0907 & 2.90 & 156 & 1809 & 2701 \\
 & 0.1239 & 1.98 & 266 & 1657 & 2429 \\
 \hline
\end{tabular}}
\caption{Comparison of the charm quark mass settings of the different collaborations through their value of $m_{D_s}$. Note that the physical value is $m_{D_s}=1968.35\pm 0.07$~MeV \cite{pdg}. $L$ and $a$ are given in fm, and the masses in units of MeV. In the last column, $M_{avg}$ stands for $M_{avg}= 1/4(m_{\eta_c} + 3m_{J\Psi}) $. }
\label{tab:comp}
\end{table}
Note that $m_{c\bar{s}}$ can vary for every LQCD simulation with a different setup. In fact, these simulations have different values for the $D_{s}$ meson mass, as shown in Table~\ref{tab:comp} where HSC refers to the Hadron Spectrum Collaboration \cite{cheungthomas}, RQCD to the work of \cite{bali} and Prelovsek et al. to \cite{lang,mohler} \footnote{Note that the values are taken from the corresponding references dividing by $a$. Also, for the data set "Prelovsek et al." we have taken the rest mass. This work the kinetic mass was tuned to the physical mass, values of the charm meson masses at rest and kinetic extracted from the dispersion relation of Eq. \ref{disprelation} are given. Here we take the rest masses for consistency with the energy levels extracted.}.

The potential $V(s)$ consistent with HQSS is then given by,
\begin{eqnarray}
 V(s)=V_{DK}(s)+V_\mathrm{ex}(s)\ .\label{eq:bare}
\end{eqnarray}
Thus the interaction considered here, Eqs.~(\ref{eq:l1}) and (\ref{eq:l2}), is derived by vector-meson exchange in $DK$ interaction supplied by the interaction with a bare state.

The interaction can also be extended to include the coupling of the $DK$ to the $D_s\eta$ channel. However, in the LQCD analyses, $D_s\eta$ interpolators are generally neglected when energies are extracted \cite{ lang,mohler, alberto,bali,cheungthomas,alexandrouc}, and thus, we omit here this channel. Indeed, the $D_s\eta\to D_s\eta$ interaction gives zero at LO using the hidden gauge lagrangian of Eq.~(\ref{eq:lag}), but not $DK\to D_s\eta$. For the two-couple channel case, naming $1$ and $2$ to the $DK$ and $D_s\eta$ channels, the term of Eq.~(\ref{eq:l1}) should be multiplied by a matrix of coefficients,
\begin{eqnarray}
 A=\left(\begin{array}{cc}1&\sqrt{2/3}\\
    \sqrt{2/3}&0
   \end{array}\right)
   \label{eq:mat1}
\end{eqnarray}
while for $V_{\mathrm{ex}}$ one multiplies by
\begin{eqnarray}
 B=\left(\begin{array}{cc}1&1/\sqrt{6}\\
    1/\sqrt{6}&1/6
   \end{array}\right)
   \label{eq:mat2}
\end{eqnarray}
and replaces $V_{c\bar{s}}^2$ in Eq.~(\ref{eq:l2}) by $V_{c\bar{s}}^iV_{c\bar{s}}^j$, with $i,j=1,2$ channels, changing also the mass of the $K$ and $D$ mesons in Eq.~(\ref{eq:couy}) by the one of the $\eta$ and $D_s$ mesons respectively for channel $2$. The constant $c$ in Eq.~(\ref{eq:couy}) is assumed to be the same in both channels, i.e., equal to the SU(3) limit. For the $\lbrace D^*K, D^*_s\eta \rbrace$ system, the equations are the same as Eqs.~(\ref{eq:l2})-(\ref{eq:mat2}), but changing $D\to D^*$, and $m_{c\bar{s}}=m_{c\bar{s}}^{1^+}$

For the analysis of the LQCD energies, the Bethe Salpeter equation in the finite volume is solved, 
\begin{equation}
 \widetilde{T}^{-1}=V^{-1}-\widetilde{G}\ ,\label{eq:bethe}
\end{equation}
where $\widetilde{T}$  is the scattering matrix in the finite volume, $V$ is given by Eq.~(\ref{eq:l1}), and $\widetilde{G}$ is the two-meson loop function in the box, 
\begin{eqnarray}
 \widetilde{G}(P^0,\vec{P}\ )=G^{DR}(P^0,\vec{P}\ )+\mathrm{lim}_{q_\mathrm{max}\to\infty}\Delta G(P_0,\vec{P},q_\mathrm{max} )\ ,\nonumber\\\label{eq:gt}
\end{eqnarray}
where $P^\mu=(P^0,\vec{P}\ )$ is the full four-momentum of the two meson system. The Mandelstam variable $s$ is related to the momentum as $s=P_0^2-\vec{P}^2=P_0^2$. The first term of Eq. (\ref{eq:gt}) is the two-meson loop function that can be evaluated in dimensional regularization as,
\begin{eqnarray}
 G^{DR}(s)&=&\frac{1}{16\pi^2}\left(\alpha+\mathrm{Log}\frac{m_1^2}{\mu^2}+\frac{m^2_2-m^2_1+s}{2s}\mathrm{Log}\frac{m^2_2}{m_1^2}\right.\nonumber\\&&\left.+\frac{p}{\sqrt{s}}\left[\mathrm{Log}\left(s-m_2^2+m_1^2+2p\sqrt{s}\right)\right.\right.\nonumber\\&&\left.-\mathrm{Log}\left(-s+m_2^2-m_1^2+2p\sqrt{s}\right)\right.\nonumber\\&&\left.+\mathrm{Log}\left(s+m_2^2-m_1^2+2p\sqrt{s}\right)\right.\nonumber\\&&\left.\left.-\mathrm{Log}\left(-s-m_2^2+m_1^2+2p\sqrt{s}\right)\right]\right)\ ,\label{eq:dim}
\end{eqnarray}
$p$ being the on-shell momentum in the center-of-mass frame, $p=\lambda^{1/2}(s,m_1^2,m_2^2)/2\sqrt{s}$.
In particular, in the center-of-mass frame (cm), where $\vec{P}=\vec{0}$, the second term in Eq. (\ref{eq:gt}), $\Delta G(P_0,\vec{P},q_\mathrm{max})$, can be written as,
\begin{eqnarray}
\Delta G=\frac{1}{L^3}\sum_{q_i}^{q_\mathrm{max}}I(\vec{q}_i)-\int_{0}^\mathrm{q_\mathrm{max}} \frac{d^3 q}{(2\pi)^3}I(\vec{q}\ )\ ,\nonumber\\\label{eq:gt1}
\end{eqnarray}
being $\vec{q}$ the momentum in the cm frame, which takes discrete values in the finite box, $\vec{q}_i=\frac{2\pi}{L}\vec{n}_i,\,\vec{n}_i\in\mathbb{Z}^3$ and $L$ the spacial extent of the box. In the continuum limit, the loop function is
\begin{eqnarray}
 G=i\int\frac{d^4q}{(2\pi)^4}\frac{1}{q^2-m_1^2+i\epsilon}\frac{1}{(P-q)^2-m_2^2+i\epsilon},
\end{eqnarray}
where $m_1$ and $m_2$ refer to masses of mesons 1 and 2 in the loop. The above formula can be evaluated, using a cutoff, $q_\mathrm{max}$,
 \begin{eqnarray}
  G^{co}=\int_0^{q_\mathrm{max}}\frac{d^3q}{(2\pi)^3}I(\vec{q}\ )\ ,
 \end{eqnarray}
where $co$ refers to the evaluation of the G function with the cutoff method and,
\begin{eqnarray}
 I(\vec{q}\ )=\frac{\omega_1(q)+\omega_2(q)}{2\omega_1(q)\omega_2(q)\left[P^2_0-(\omega_1(q)+\omega_2(q))^2+i\epsilon\right]}\ ,
\end{eqnarray}
with $\omega_i=\sqrt{q^{2}+m_i^2}$, and $q=|\vec{q}\ |$.

For the moving frame and partial wave decomposition of the scattering amplitude in the finite volume, we follow the procedure described in \cite{doringframe}.  The two-meson loop function in the finite volume, $\widetilde{G}$ in Eq.~(\ref{eq:gt}) projected in partial waves, is a matrix with non-diagonal elements due to the partial weave mixing. Its elements are,
\begin{eqnarray}
    \widetilde{G}_{lm,l'm'}=\frac{4\pi}{L^3}\sum_{n_i}^{q_\mathrm{max}}\frac{\sqrt{s}}{P_0}\left(\frac{q}{q^\mathrm{on}}\right)^kY_{lm}^*(\hat{q})Y_{l'm'}(\hat{q})I(q)\ ,\nonumber\label{eq:gdis}\\
\end{eqnarray}
with $k=0,1$ for $l+l'=$ even, odd~\cite{doringframe}. The relevant equation to evaluate the energy levels in the finite volume, is,
\begin{eqnarray}
    \mathrm{det}(\delta_{ll'}\delta_{mm'}-V_l\tilde{G}_{lm,l'm'})=0\label{eq:end}
\end{eqnarray}
$V_l$ is the interaction projected in $l$-wave. Here, we include only $S$ and $P$ partial waves of the interaction given in Eq.~(\ref{eq:l1}). The irreducible representations which appear for a given external momenta are discussed in section II.E of \cite{doringframe}. 

In the following section we show the results of the analysis of the LQCD energy levels of HSC \cite{cheungthomas} for $m_\pi=239$ (HSC239) and $391$~MeV (HSC391).  We will compare with previous analyses of LQCD data from other collaborations \cite{liuorginos,lang,mohler,bali} done in \cite{liuorginos,alberto,pedro}. A combined fit of the most recent LQCD energy levels \cite{lang,mohler,bali,cheungthomas} will be performed, and the quark mass dependence of the $D_{s0}^*(2317)$ extracted from this fit. Here, we use the dimensional regularization formula of Eq.~(\ref{eq:dim}) to evaluate the two meson loop function in the finite volume of Eqs.~(\ref{eq:gt}) and~(\ref{eq:gt1}). If we consider only the meson-meson interaction given by Eq.~(\ref{eq:l1}), the only free parameter is the subtraction constant, $\alpha$ (we fix $\mu=1000$~MeV \cite{ollermeissner}). However, there are energy levels for two different pion masses in the case of \cite{cheungthomas}. The subtraction constant can be different for the energy levels of the light and heavy pion mass. Thus, we assume a first order taylor expansion of the subtraction constant in the pion mass, $\alpha=\alpha_1+\alpha_2 m_\pi^2$. The effect of $\alpha_2$ can be understood as to reabsorb possibly relevant higher order contributions in the potential of Eqs.~(\ref{eq:l1}) and (\ref{eq:bare}) in the Bethe-Salpeter equation. Then, we end up with two free parameters for every data set when it is analyzed separately. On the other hand, we also compare the  result using the interaction of Eq.~(\ref{eq:l1}) with the one adding a bare component, Eqs.~(\ref{eq:l2})-(\ref{eq:bare}). In this case we have a couple of two more free parameters, $c$, and $m_{c\bar{s}}$. The mass $m_{c\bar{s}}$ is then tuned for every LQCD collaboration according to Table~\ref{tab:comp}. 

The scattering amplitude in the infinite volume is,
\begin{eqnarray}
 T^{-1}(s)=V^{-1}(s)-G(s)\ .\label{eq:inf}
\end{eqnarray}
For the energies in the finite volume, $V^{-1}(E^0)=\widetilde{G}(E^0)$, and Eq.~(\ref{eq:inf}) becomes equivalent to the Lüscher formalism~\cite{Luscher1,Luscher2} up to
exponentially suppressed corrections~\cite{doringru}. Near to a pole, $s_0$, the scattering amplitude behaves as,
\begin{eqnarray}
 T\simeq\frac{g_ig_j}{s-s_0}\label{eq:cou}
\end{eqnarray}

We will study the quark mass dependence of the pole position and its \textit{compositeness} \cite{Weinbergcom}. For that, we evaluate the probability of finding the $DK$ molecular component in the wave function~\cite{gamermanncou,Hyodo,ollerzhi},
\begin{equation}
 P_i=- g^2_i\left(\frac{\partial G}{\partial s}\right)_{s=s_0}\ .
\end{equation}
where $i$ refers to the $DK$ channel. In general, one has the sum rule~\cite{gamermanncou,Hyodo,ollerzhi},
\begin{eqnarray}
 -\sum_ig_i^2\frac{\partial G_i}{\partial s}-\sum_{i,j}g_ig_jG_i\frac{\partial V_{ij}}{\partial s}G_j=1\ ,
\end{eqnarray}
where the first term is associated to the \textit{compositeness} ($1-Z$), and the second one to the \textit{elementariness} ($Z$).

Finally, the scattering length and effective range can be evaluated using the effective range approximation, which  reads
\begin{equation}
    p\mathrm{cot}\,\delta=\frac{1}{a}+\frac{1}{2}r_0 p^2\ .
\end{equation}

\section{Results}
\subsection{Energy levels and quark mass dependence}
We analyze the energy levels of \cite{cheungthomas} for $DK$ scattering in $I=0$, with the interaction given by Eqs.~(\ref{eq:l1}) and (\ref{eq:bare}), by solving Eq.~(\ref{eq:end}), with $\tilde{G}_{l'm',lm}$ given by Eq.~(\ref{eq:gdis}).
As fitting parameters, we have, $c$, $m_{c\bar{s}}$, $\alpha_1$ and $\alpha_2$ for the subtraction constants. The $\chi^2$ is given by
\begin{eqnarray}
    \chi^2=\Delta E^TC^{-1}\Delta E
\end{eqnarray}
where $\Delta E_i=E_i-E^0_i$, with $E^0_i$ the lattice energy $i$, and $C$ de covariance matrix. The covariance matrix is taken from the analysis of \cite{cheungthomas}.

We consider two situations, one where the bare $c\bar{s}$ component is not included, i.e., $V_\mathrm{ex}=0$, and other with $V_\mathrm{ex}\neq 0$. We also compare the results from the fits of energy levels of \cite{cheungthomas} with previous LQCD analyses \cite{pedro,alberto}, for the LQCD data of \cite{bali,mohler,lang}, and with the quark mass dependence obtained for the scattering lengths with previous works \cite{liuorginos}.

\begin{figure}
    \centering
    \begin{minipage}{0.461\textwidth}
     \centering
     \includegraphics[width=1.0\linewidth]{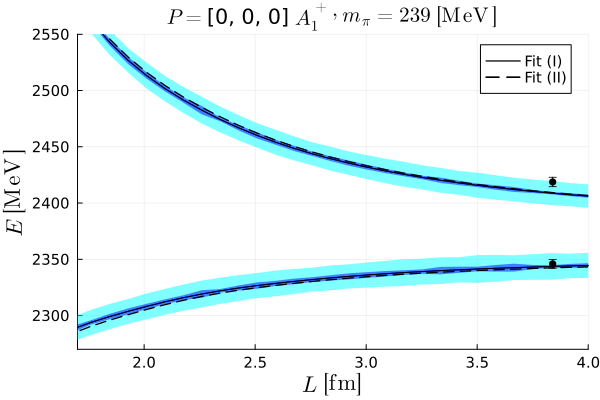}
   \end{minipage}
    \centering
    \begin{minipage}{0.461\textwidth}
     \centering
     \includegraphics[width=1.0\linewidth]{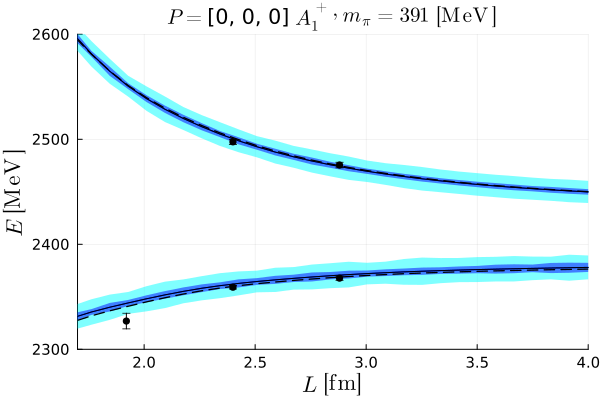}
   \end{minipage}
   \centering
    \begin{minipage}{0.461\textwidth}
     \centering
     \includegraphics[width=1.0\linewidth]{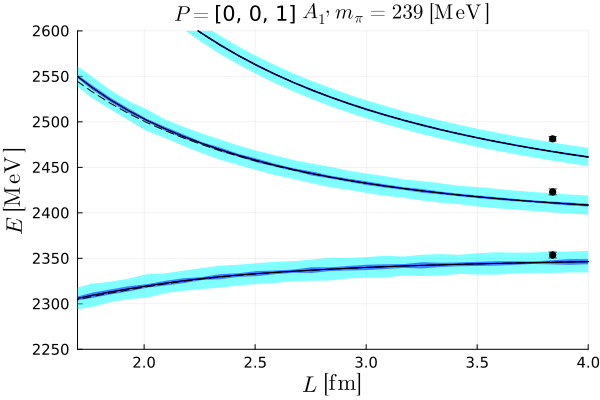}
   \end{minipage}
   \centering
    \begin{minipage}{0.461\textwidth}
     \centering
     \includegraphics[width=1.0\linewidth]{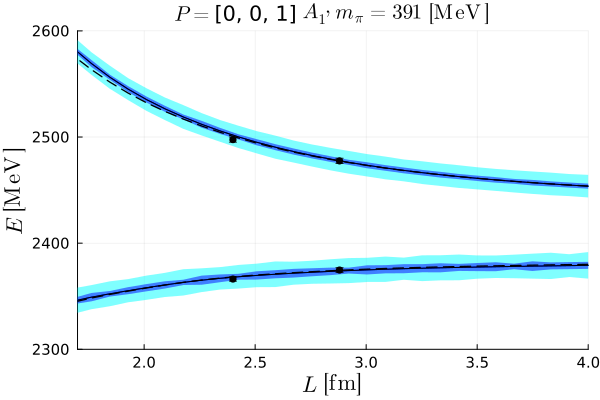}
   \end{minipage}
    \caption{Result from the fit for the dependence with the spatial extent (L) of the energy levels of HSC for $P=(0,0,0)$ and $P=(0,0,1)2\pi/L$. Continuous lines are for Fit I, and dashed line for Fit II. The result for other momenta can be found in the Appendix, Sec.~\ref{sec:boosts}, Fig.~\ref{fig:energylvlshsc} and Fig.~\ref{fig:energylvlshsc1}.}
    \label{fig:niveleshsc}
\end{figure}

First, we show the results for the LQCD energy levels, Fit I ($V_\mathrm{ex}=0$), and II ($V_\mathrm{ex}\neq 0$) in Fig.~\ref{fig:niveleshsc} for $\vec{P}=\vec{0}$ and $\vec{P}=(0,0,1)2\pi/L$. See also Figs.~\ref{fig:energylvlshsc} and \ref{fig:energylvlshsc1} in the Appendix, sec.~\ref{sec:boosts}, for other boosts. The energy levels are well described in both models, as shown in these figures. These barely change between Fit I and II.

The parameters obtained in both fits are shown in Table~\ref{tab:par1}.
\begin{table}[htb]
\setlength{\tabcolsep}{0.3em}
{\renewcommand{\arraystretch}{1.6}
\centering

    \begin{tabular}{|c|c|c|c|c|c|}
    \hline
    &$c$&${m}_{c\bar{s}}^{L}$& ${m}_{c\bar{s}}^{H}$&$\alpha_1$&$\alpha_2[(m_\pi^{0})^{-2}]$\\
    \toprule
     Fit I & - & - & - & $-1.86(1)$ & $-0.041(12)$   \\
     Fit II & $1.02(4)$ & $2508(61)$ & $2445(43)$ & $-1.34(6)$ & $0.000(2)$ \\
     \hline
    \end{tabular}
    \caption{Values of the parameters obtained in the fit of the HSC data sets, in both fits, I and II, as explained in the text. $\alpha_2$ is given in units of $(m_\pi^{0})^{-2}$, with $m_\pi^{0}$ the physical pion mass. The superscript $L$ means light pion mass and $H$ stands for the heavier in Table~\ref{tab:comp}.}
    \label{tab:par1}}
\end{table}
The magnitudes of the parameters $\alpha_1$, $\alpha_2$ are significantly reduced when the bare component is included. Indeed, the pion mass dependence of the subtraction constant is now reabsorbed in the new term of the interaction given by the coupling of the $c\bar{s}$ bare component to $DK$, so that, we obtain $\alpha_2=0$ in Fit II. The fact that the energy levels are well described in both cases, means that the $\alpha_1$, $\alpha_2$ parameters are able to absorb the coupling to the genuine component in the simple model of Fit I. We find only one pole in this sector. The pole position and coupling of the $D_{s0}^*(2317)$ to $DK$ are evaluated by using Eqs.~(\ref{eq:inf}) and~(\ref{eq:cou}). The results for the pole position, couplings, and compositeness ($1-Z$), are given in Tables~\ref{tab:res1} and~\ref{tab:res1a}, for the two different pion masses studied and the physical pion mass.
\begin{table}[htb]
\setlength{\tabcolsep}{0.4em}
{\renewcommand{\arraystretch}{1.6}
\centering

    \begin{tabular}{|c|c|c|c|c|}
    \hline

    $m_\pi$&$m_{\eta_c}$&$\sqrt{s_0}$&$g$ [GeV]& $1-Z$\\
    \toprule
    $138$ & $2984$ & $2323(6)$ & $10.42(10)$ & $0.733(9)$ \\
     $239$ & $2986^*$ & $2350(12)$ & $10.34(12)$ & $0.723(13)$ \\
    $391$ & $2964^*$ & $2384(14)$ & $11.13(11)$ & $0.659(17)$ \\
     \hline
    \end{tabular}
    \caption{Fit I: Values of the pole mass, the coupling of the state to $DK$, and its \textit{compositeness} for the different values of the pion and the $\eta_c$ meson masses. The masses and pole position, $\sqrt{s_0}$, are given in units of MeV.}
    \label{tab:res1}}
\end{table}
\footnote{The values of the $\eta_c$ mass for the light and heavy pion mass are obtained from $(a_tm_{\eta_c})^2=0.2412,0.2735$, respectively \cite{Cheungligero,Liumingpesado}.}
\begin{table}[htb]
\setlength{\tabcolsep}{0.4em}
{\renewcommand{\arraystretch}{1.6}
\centering
    \begin{tabular}{|c|c|c|c|c|}
    \hline

    $m_\pi$&$m_{\eta_c}$&$\sqrt{s_0}$&$g$ [GeV]&$1-Z$\\
    \toprule
    $138$ & $2984$ & $2322(22)$ & $10.77(13)$ & $0.770(61)$ \\
     $239$ & $2986^*$& $2349(27)$ & $10.69(15)$ & $0.761(71)$ \\
    $391$ & $2964^*$& $2382(36)$ & $11.57(13)$ & $0.696(76)$\\
     \hline
    \end{tabular}
    \caption{Same as Table~\ref{tab:res1} but for Fit II.}
    \label{tab:res1a}}
\end{table}
The result obtained for the physical point is in very good agreement with the experiment. 

\begin{figure}
    \centering
    \begin{minipage}{0.461\textwidth}
     \centering
     \includegraphics[width=1.0\linewidth]{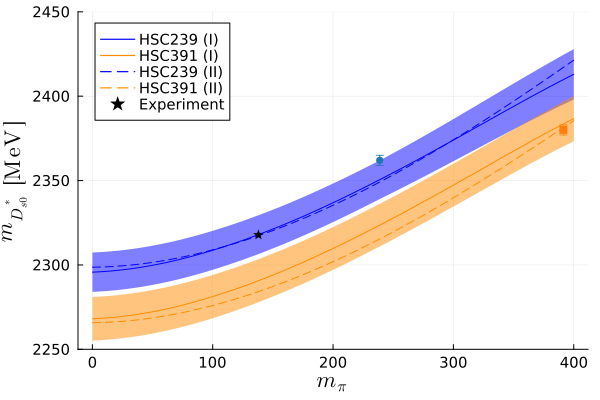}
   \end{minipage}
   \begin{minipage}{0.461\textwidth}
     \centering
     \includegraphics[width=1.0\linewidth]{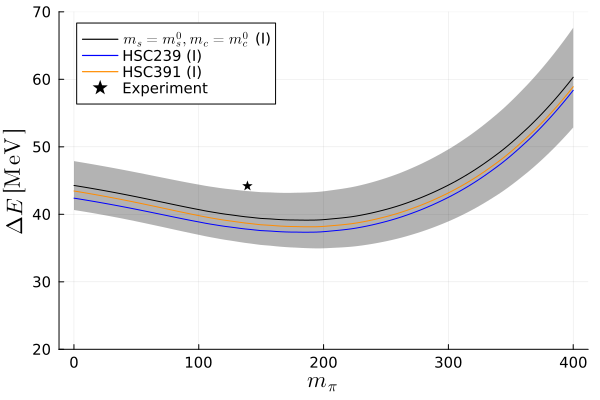}
   \end{minipage}
   \begin{minipage}{0.461\textwidth}
     \centering
     \includegraphics[width=1.0\linewidth]{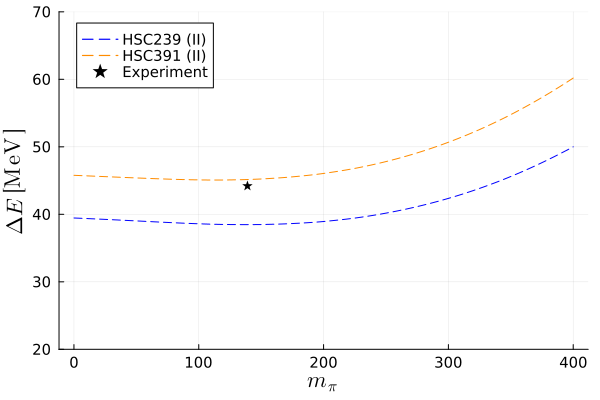}
   \end{minipage}
   \begin{minipage}{0.461\textwidth}
     \centering
     \includegraphics[width=1.0\linewidth]{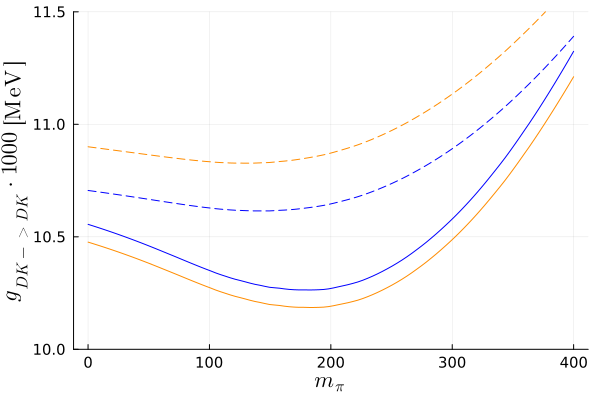}
   \end{minipage}
    \caption{Pole mass, binding energy and coupling of the state to the $DK$ component as a function of the pion mass for the two HSC data sets. Continuous lines indicate Fit I, while the dashed lines are for Fit II. The black lines are showing the extrapolation to $m_c=m_c^0$, and $m_s=m_s^0$.}
    \label{fig:masscou}
\end{figure}

We can explore further the quark mass dependence of the $D_{s0}^*(2317)$ properties by taking as input the study of the low-lying charmed meson masses of~\cite{gil}, where this dependence is studied for the $D^{(*)}_{(s)}$ mesons from an analysis of LQCD data with HH$\chi$PT. The results are shown in Fig.~\ref{fig:masscou}. In the top panel, the pole position and the binding energy is shown. The analysis of~\cite{gil} takes into account the different charm quark mass of the HSC LQCD data sets, which are also slightly different than the physical charm quark mass. The binding energies obtained in Fits I and II are plotted in the middle panels, where we also show the extrapolation to $m_c=m_c^0$ and $m_s=m_s^0$ \footnote{The superscript $0$ stands for physical value.} as a continuous black line for Fit I. The error band englobes the physical point.

\begin{figure}
    \centering
    \begin{minipage}{0.461\textwidth}
     \centering
     \includegraphics[width=1.0\linewidth]{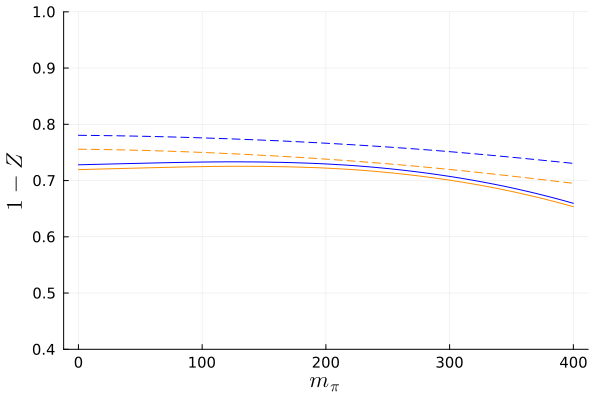}
   \end{minipage}
    \caption{Evolution of the \textit{compositeness}, $1-Z$, with the pion mass, for the two different HSC data sets as a result of Fits I and II.}
    \label{fig:1mz}
\end{figure}

The coupling of the state to the $DK$ component is shown in the lower panel of Fig.~\ref{fig:masscou}. We see that the coupling has a value between $(10 - 12) \cdot 10^3$~MeV in the range of pion masses studied here. With this, we can evaluate the compositeness, shown in Fig.~\ref{fig:1mz}. One can conclude that in the range of pion masses studied, this is almost constant around $0.7-0.8$, slightly decreasing when the state is moving further from the threshold and higher in Fit II. These values are consistent with previous works~\cite{liuorginos,alberto,pedro,Yang:2021tvc}.

\begin{figure}
    \centering
    \begin{minipage}{0.461\textwidth}
     \centering
     \includegraphics[width=1.0\linewidth]{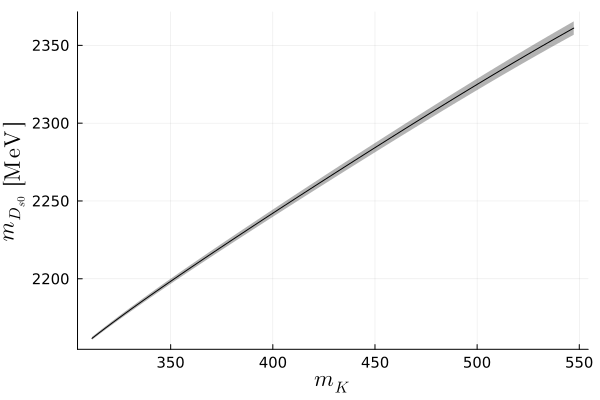}
   \end{minipage}
    \caption{Evolution of the $D_{s0}^*(2317)$ pole mass with the kaon mass, fixing $m_\pi$ to its physical value.}
    \label{fig:ds0mk}
\end{figure}

The type of pion mass dependence obtained here for the $D_{s0}^*(2317)$ pole is quadratic for very light pion masses, as in Ref.~\cite{Clevenmass}. However, for larger pion masses it clearly becomes linear, see Fig.~\ref{fig:masscou}. Thus, the prediction of this dependence here is milder than in \cite{Clevenmass}. We also plot the pole evolution with the mass of the kaon for $m_\pi=m_\pi^0$, the physical value, in Fig.~\ref{fig:ds0mk}. We obtain a linear dependence with slope near $1$. This is in agreement with~\cite{Clevenmass}.

\begin{figure}
    \centering
    \begin{minipage}{0.461\textwidth}
     \centering
     \includegraphics[width=1.0\linewidth]{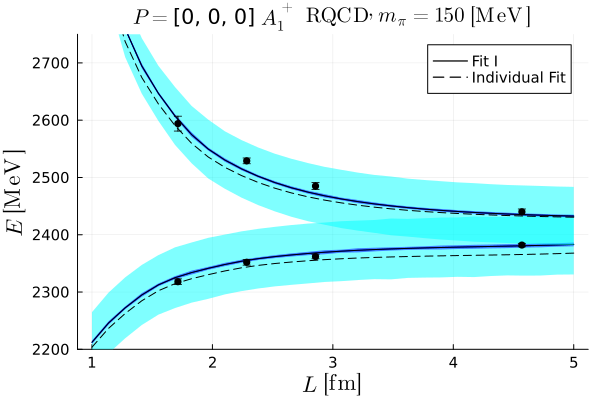}
   \end{minipage}
    \begin{minipage}{0.461\textwidth}
     \centering
     \includegraphics[width=1.0\linewidth]{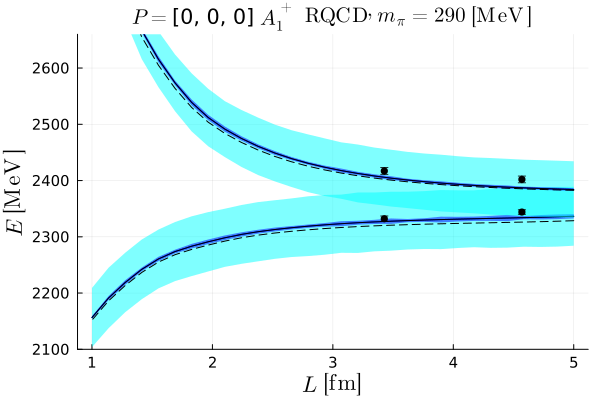}
   \end{minipage}
   \begin{minipage}{0.461\textwidth}
     \centering
     \includegraphics[width=1.0\linewidth]{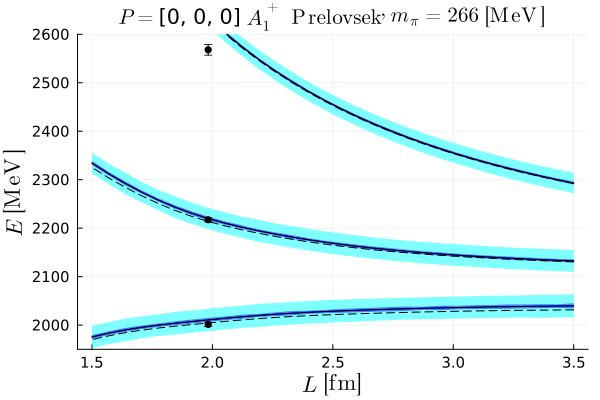}
   \end{minipage}
       \caption{Result from the fit for the dependence with the spatial extent (L) of the $DK$ energy levels obtained using the parameters $\alpha_1,a_2$ (Fit I) in Table~\ref{tab:par1}, for $m_\pi=150, 290$, \cite{bali} (first and second panels from the top) and $m_\pi=266$ MeV \cite{lang,mohler} (third panel), compared to the result from individual fits of~\cite{bali} and \cite{lang,mohler}.}
    \label{fig:basa}
\end{figure}

\begin{figure}
    \centering
    \begin{minipage}{0.461\textwidth}
     \centering
     \includegraphics[width=1.0\linewidth]{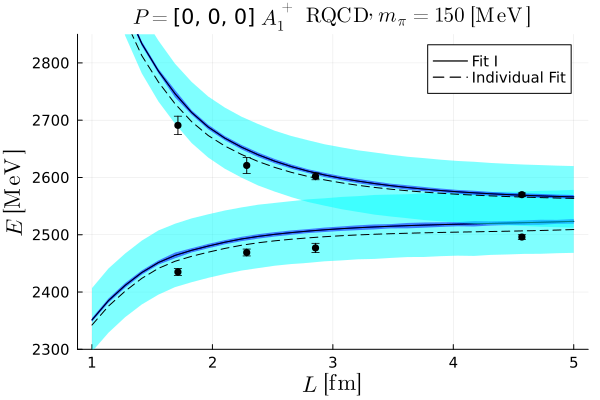}
   \end{minipage}
   \begin{minipage}{0.461\textwidth}
     \centering
     \includegraphics[width=1.0\linewidth]{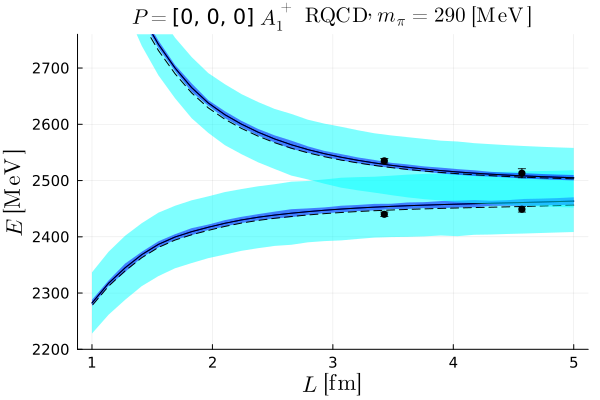}
   \end{minipage}
   \begin{minipage}{0.461\textwidth}
     \centering
     \includegraphics[width=1.0\linewidth]{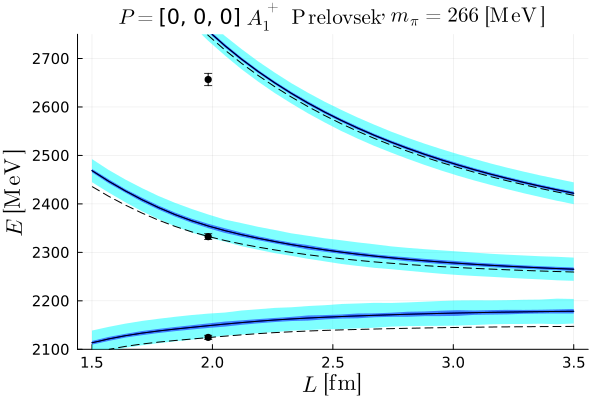}
   \end{minipage}
    \caption{Same as Fig.~\ref{fig:basa} but with $D^*K$ energy levels.}
    \label{fig:basa_ds1}
\end{figure}

Next, we compare with the results of previous works. Looking at the energy levels of \cite{bali} which were analyzed in \cite{pedro}, and \cite{lang,mohler}, analyzed in \cite{alberto}, we can ask ourselves how good is the prediction of the energy levels of \cite{bali,lang,mohler} using the parameters given in Table~\ref{tab:par1}, where we have selected Fit I. The prediction of the energy levels using $\alpha_1,a_2$ for the pion masses of \cite{bali,pedro} ($150$ and $290$ MeV), and ensemble (1) of \cite{lang,mohler,alberto} ($266$ MeV), are shown in Fig.~\ref{fig:basa} and ~\ref{fig:basa_ds1} with black continuous lines\footnote{In \cite{lang,mohler} energy levels are also given for another ensemble, called ensemble (2), with $m_\pi=156$ MeV. However, we do not include this ensemble here because, even though we obtain a good agreement for the energy levels in our fit, the mass of the $D_{s0}^*(2317)$ obtained differs significantly from the one given in \cite{lang,mohler}.}. The agreement is indeed quite good. We also conduct a fit of those, being the result shown in the same figure (of Fit I type) with dashed lines. The energy levels barely change for the RQCD Collaboration (first and second panels are for $m_\pi=150$ and $290$ MeV, respectively), while for the [Prelovsek et al.] data set, these change only slightly (third panel is for  $266$ MeV), being still inside the error band.

\begin{figure}[h!]
    \centering
   \begin{minipage}{0.44\textwidth}
     \centering
     \includegraphics[width=1.0\linewidth]{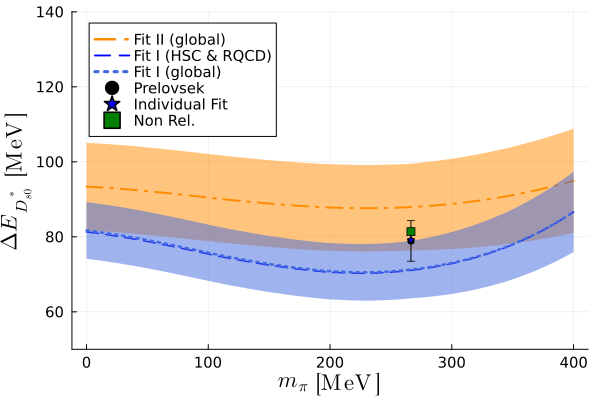}
   \end{minipage}
   \begin{minipage}{0.44\textwidth}
     \centering
     \includegraphics[width=1.0\linewidth]{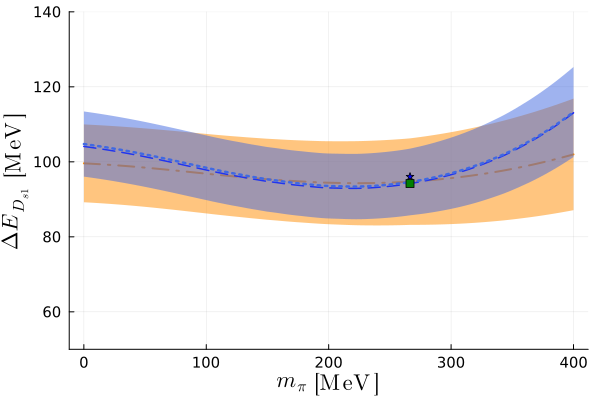}
   \end{minipage}
   \begin{minipage}{0.44\textwidth}
     \centering
     \includegraphics[width=1.0\linewidth]{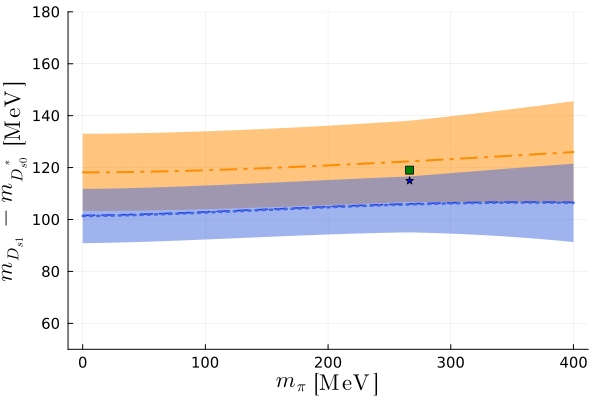}
   \end{minipage}
\caption{Pion mass dependence of the $D_{s0}^*(2317)$ and $D_{s1}(2460)$ binding energies and splitting. Black point correspond to the result of \cite{lang,mohler}, the results of our individual fits are shown with a star (relativistic approach) and green square (non relativistic dispersion relation).}
    \label{fig:MOHLER}
\end{figure}

\begin{figure}
    \centering
    \begin{minipage}{0.45\textwidth}
     \centering
     \includegraphics[width=1.0\linewidth]{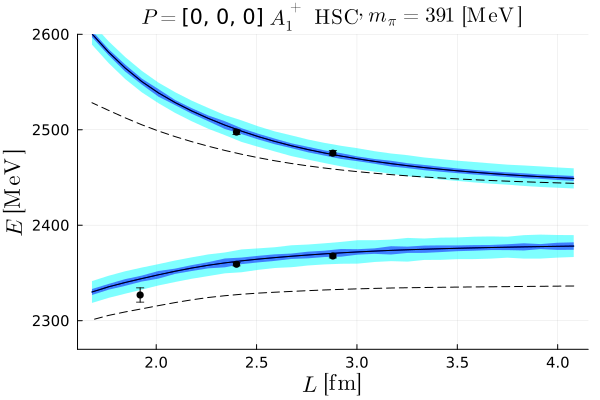}
   \end{minipage}
   \begin{minipage}{0.45\textwidth}
     \centering
     \includegraphics[width=1.0\linewidth]{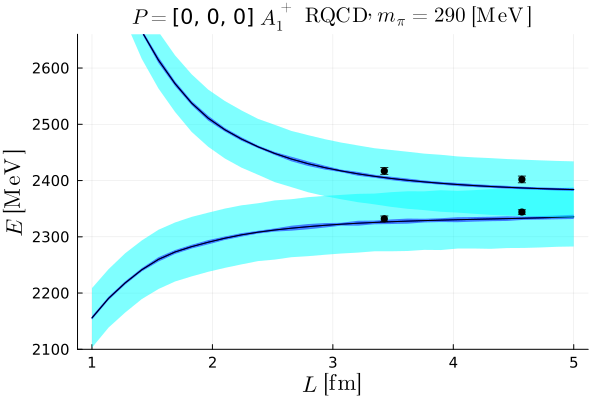}
   \end{minipage} 
   \begin{minipage}{0.45\textwidth}
     \centering
    \includegraphics[width=1.0\linewidth]{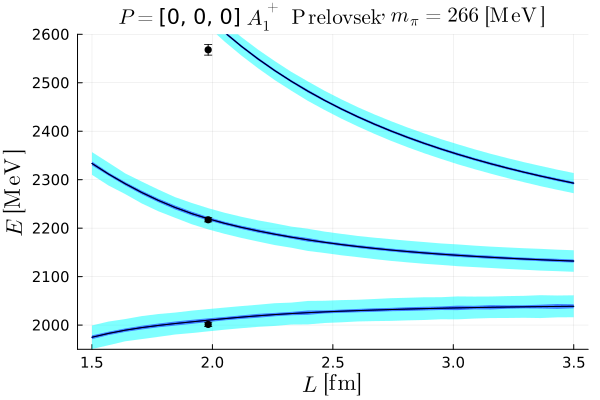}
   \end{minipage} 
    \caption{Fit I: Result from the global fit for the dependence with the spatial extent (L) of some energy levels of \cite{cheungthomas} (upper panel), \cite{bali} (middle), and \cite{lang,mohler} (lower panel). In the upper panel, the effect of the $D_s\eta$ channel is shown in dashed lines.}
    \label{fig:comb}
\end{figure}

These results suggest that it may be possible to perform a global fit with the energy levels of HSC~\cite{cheungthomas}, and from \cite{bali,mohler,lang}. Note that in \cite{bali,mohler,lang} the $D_{s1}(2460)$ has also been simulated through the $D^*K$ scattering. We also include here the energy levels corresponding to the $D_{s1}(2460)$. Thus, we perform a combined fit of the one-channel $DK$ and $D^*K$ energy levels. In Fig.~\ref{fig:comb} we show the result for the energy levels at $P=0$, for HSC, RQCD, and [Prelovsek et al.] in one channel. The two first energy levels for the different data sets are reproduced well. Note that in \cite{lang,mohler} a non relativistic dispersion relation based on the so-call Fermilab approach is used:
\begin{equation}\label{disprelation}
    E_{D(D^*)}(\vec{p})=M_1+\frac{\vec{p}^2}{2M_2} - \frac{(\vec{p}^2)^2}{8M_4^3},
\end{equation}
where $M_1=1561$ MeV, $M_2=1763$ MeV and $M_4=1763$ MeV for the $D$ meson and $M_1=1693$ MeV, $M_2=2018$ MeV and $M_4=2110$ MeV for the $D^*$ meson. Note that, in Eq. (\ref{disprelation}), the masses $M_1$, $M_2$ and $M_4$ do not correspond to the same masses obtained in lattice approaches with relativistic charm quarks. However, they should become the same in the continum limit. More discussion on this can be found in \cite{El-Khadra:1996wdx}.
We have checked the result of Fit I  using this dispersion relation, as done in the reanalysis of Ref. \cite{alberto}, and we have found that, a result very close to the one with a relativistic dispersion relation is obtained. Also, ensemble (1) included in the global fit has a negligible effect in the final result of the global fit. Indeed, we obtain the same central value and error band if we do not include this ensemble, as shown in Fig.~\ref{fig:MOHLER}. However, as mentioned before, the two first energy levels are well described, see Fig.~\ref{fig:comb}. 

We would like to make a consideration. In Table~\ref{tab:res1} we showed the result of the $D_{s0}^*(2317)$ pole for the physical point for the analysis of the energy levels of the HSC. One can see that the mass of the $D_{s0}^*(2317)$ is compatible with the experimental point. However, at the physical point, there is one more channel near the $DK$ threshold, the $D_s\eta$ channel. To understand the role of this channel at the physical point, we predict, with the result of the global fit, the effect of the $D_s\eta$ channel for the HSC data. This is shown in Fig.~\ref{fig:comb} with dashed lines. The coupling to the $D_s\eta$ channel tends to lower down the energy levels suggesting an attractive interaction. This can be understood as one can eliminate the $D_s\eta$ channel introducing an effective potential for the case of two channels as, $V'_{11}=V_{11}+V_{12}^2G_2$, with $2:D_s\eta$, where $G_2$ is negative \cite{Acetioset}.

The set of parameters obtained in the global fit are collected in Tables~\ref{tab:par2} and ~\ref{tab:par2x}. We obtain a reasonable value for the $D_{s1}-D_{s0}^*$ mass splitting by letting the parameter $\alpha_1$ to sligthly vary for the $D_{s1}$ state, that we call $\alpha_1^{1^+}$. As can be seen, these parameters barely change respect to the ones in Table~\ref{tab:par1}. 
\begin{table}[htb]
\setlength{\tabcolsep}{0.4em}
{\renewcommand{\arraystretch}{1.6}
\centering

    \begin{tabular}{|c|c|c|c|c|}
    \hline
    &$c$&$\alpha_1^{0^+}$&$\alpha_1^{1^+}$&$\alpha_2[(m_\pi^{0})^{-2}]$\\
    \toprule
     Fit I & - & $-1.87(1)$ & $-2.05(2)$ & $-0.040(2)$   \\
     Fit II & $1.05(1)$ & $-1.34(1)$ & $-1.44(1)$ & $-0.002(1)$    \\
     \hline
    \end{tabular}
    \caption{Common parameters among all collaborations obtained in the combined LQCD anaylis (global fit), for Fits I and II.}
    \label{tab:par2}}
\end{table}

\begin{table}[htb]
\setlength{\tabcolsep}{0.4em}
{\renewcommand{\arraystretch}{1.6}
\centering

    \begin{tabular}{|c|c|c|}
    \hline
    &${m}_{c\bar{s}}^{0^+}$& ${m}_{c\bar{s}}^{1^+}$ \\
    \toprule
     HSC239 & $2552(26)$ & -    \\
     HSC391 & $2476(25)$ & -  \\
     RQCD & $2475(25)$ & $2553(25)$  \\
     Prelovsek & $2277(23)$ & $2356(23)$  \\
     \hline
    \end{tabular}
    \caption{Parameters of the ${m}_{c\bar{s}}$ CQM bare masses obtained in the combined LQCD anaylis (global fit) for Fit II.}
    \label{tab:par2x}}
\end{table}

\begin{figure*}
   \begin{minipage}{0.46\textwidth}
     \centering
     \includegraphics[width=1\linewidth]{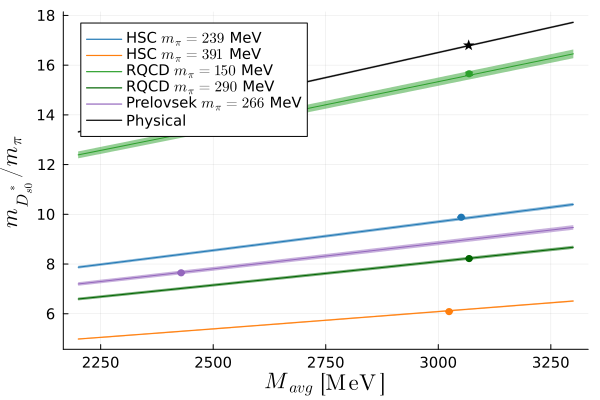}
   \end{minipage}
   \begin{minipage}{0.46\textwidth}
     \centering
     \includegraphics[width=1\linewidth]{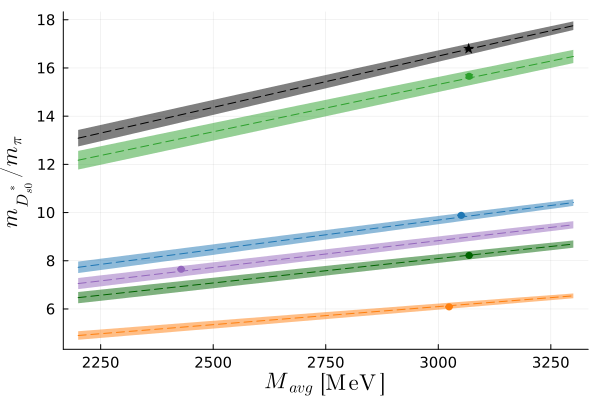}
   \end{minipage} 
   \begin{minipage}{0.46\textwidth}
     \centering
     \includegraphics[width=1\linewidth]{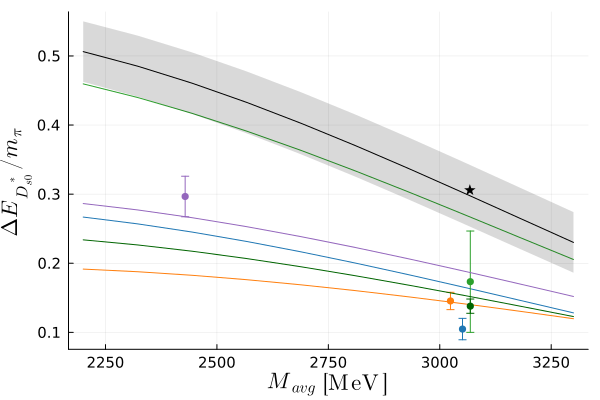}
   \end{minipage} 
   \begin{minipage}{0.46\textwidth}
     \centering
     \includegraphics[width=1\linewidth]{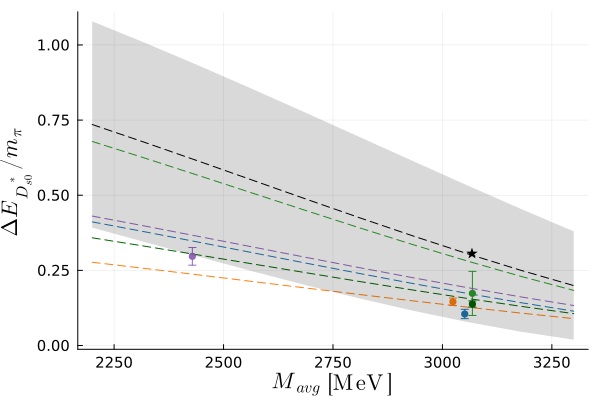}
   \end{minipage} 
   \begin{minipage}{0.46\textwidth}
     \centering
     \includegraphics[width=1\linewidth]{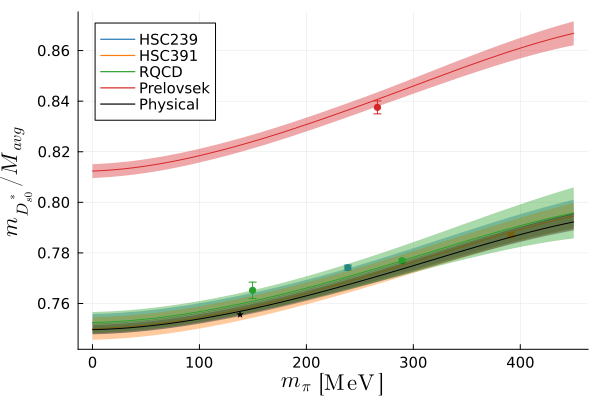}
   \end{minipage}
   \begin{minipage}{0.46\textwidth}
     \centering
     \includegraphics[width=1\linewidth]{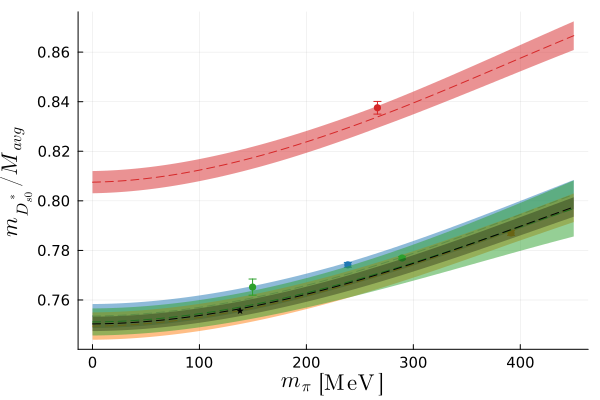}
   \end{minipage} 
   \begin{minipage}{0.46\textwidth}
     \centering
     \includegraphics[width=1\linewidth]{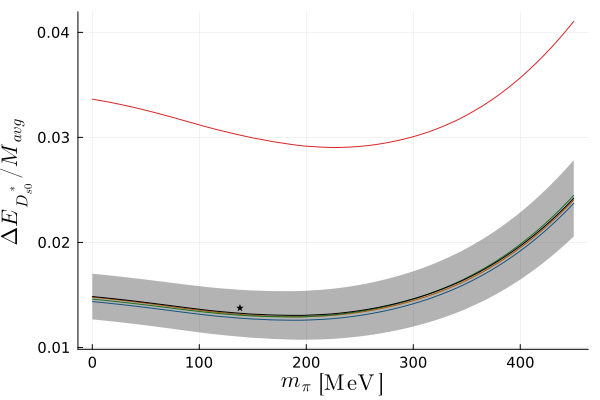}
   \end{minipage} 
   \begin{minipage}{0.46\textwidth}
     \centering
     \includegraphics[width=1\linewidth]{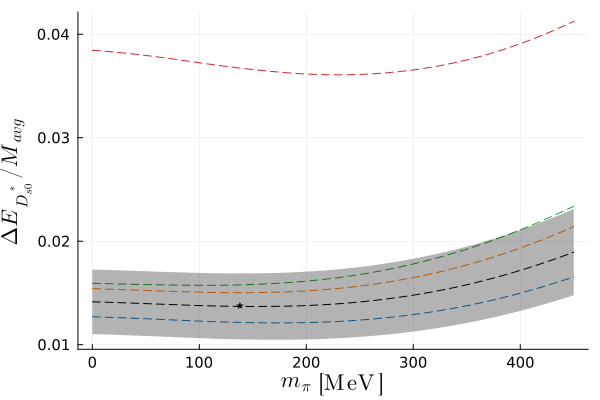}
   \end{minipage} 
   \caption{Result of the global fits for the quark mass dependence of the pole mass and binding energy for Fit I (left panels) and Fit II (right panels).}
   \label{fig:masscout1}
\end{figure*}

\begin{figure}[h!]
    \centering
    \begin{minipage}{0.44\textwidth}
     \centering
     \includegraphics[width=1.0\linewidth]{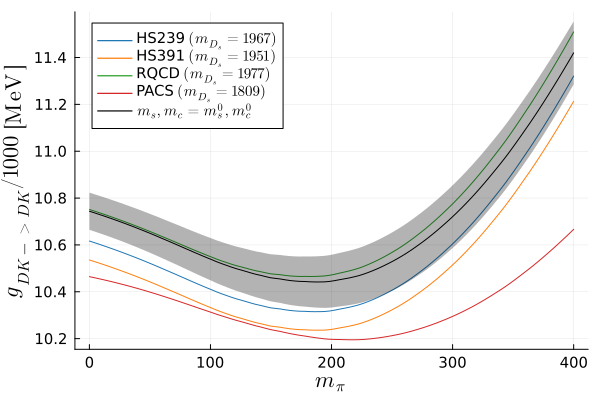}
   \end{minipage}
    \begin{minipage}{0.44\textwidth}
     \centering
     \includegraphics[width=1.0\linewidth]{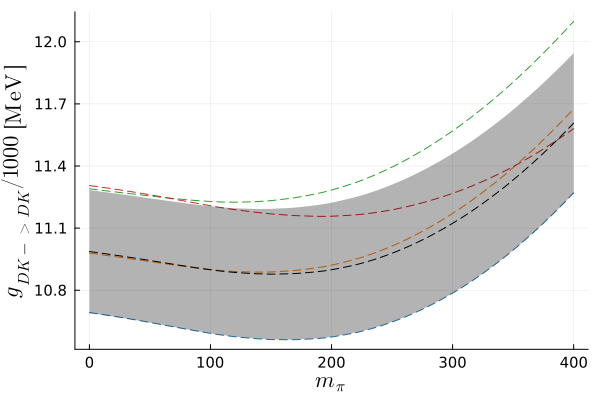}
   \end{minipage}
    \caption{Result of the global fit (Fit I top panel, Fit II bottom panel) for the coupling as a function of the pion mass.}
    \label{fig:coupling1}
\end{figure}

With these parameters and the input of the quark mass dependence of the $D(D^*)$ meson~\cite{gil}, we evaluate the dependence of the $m_{D_{s0}^*}$, the binding energy and its coupling to the $DK$ channel in the one channel case. The results are given in Figs.~\ref{fig:masscout1} and ~\ref{fig:coupling1}. In Fig.~\ref{fig:masscout1}, the ratios, $m_{D_{s0}^*}/m_\pi$, $\Delta E_{D_{s0}^*}/m_\pi$, as a function of $M_{avg}$, and $m_{D_{s0}^*}/M_{avg}$, $\Delta E_{D_{s0}^*}/M_{avg}$, as a function of $m_\pi$, are shown from top to bottom, respectively \footnote{The dependence with $M_{avg}$ is taken from the result of the analysis of \cite{gil}, where, the parameter depending on the heavy quark mass in the charm meson masses, $m_H$, satisfies closely for all lattice data sets studied, $m_H=0.635 M_{avg}$.}. The mass of the $D_{s0}^*(2317)$ increases with the charm quark mass for all the collaborations. Remarkably, the binding energy (with respect to the respective thresholds of the LQCD collaborations) becomes larger for lower charm quark masses, suggesting that the interaction gets less attractive when the charm quark mass increases, as depicted in the second panels from the top. Note that, since the binding energy, $\Delta E_{D_{s0}^*}=m_K+m_D-m_{D_{s0}^*}$, this means that the pole of the $D_{{s_0}}^*$ is moving faster than the threshold when the charm quark mass ($M_{avg}$) increases. For Fit I, the error band obtained for the physical point is narrow and this effect is clear. However, as shown in Fig.~\ref{fig:masscout1}, this error band is very broad in the case of Fit II\footnote{Only the error band for the physical pion mass is shown for clarity.}. This is because the number of free parameters have increased. The dependence with the charm quark mass is not unambiguously determined in this case. Therefore, more lattice data for lower than the physical charm quark mass are needed to test this prediction. Nevertheless, the behaviour with the light quark mass is opposite. For pion masses larger than $200$ MeV, the binding energy clearly increases, becoming the interaction more attractive. 

\begin{figure}[h!]
    \centering
    \begin{minipage}{0.45\textwidth}
     \centering
     \includegraphics[width=1.0\linewidth]{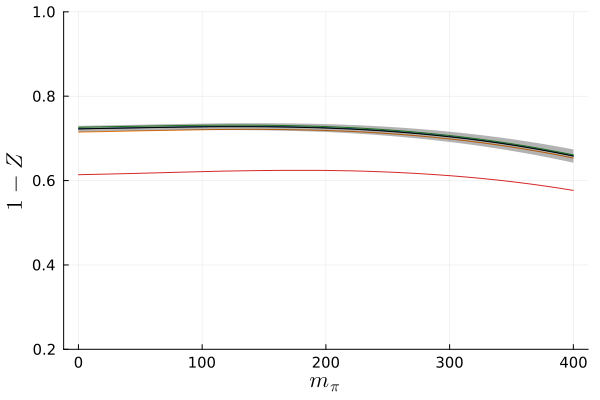}
   \end{minipage}
   \begin{minipage}{0.45\textwidth}
     \centering
     \includegraphics[width=1.0\linewidth]{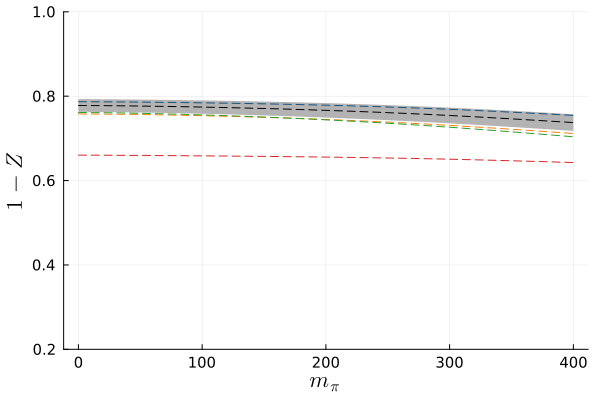}
   \end{minipage}
    \caption{Result of the global fit (Fit I top panel, Fit II bottom panel) for the \textit{compositeness} as a function of the pion mass (same legend than in Fig.~\ref{fig:masscout1} bottom).}
    \label{fig:1mz1}
\end{figure}

\begin{figure*}
    \centering
    \begin{minipage}{0.44\textwidth}
     \centering
     \includegraphics[width=1.0\linewidth]{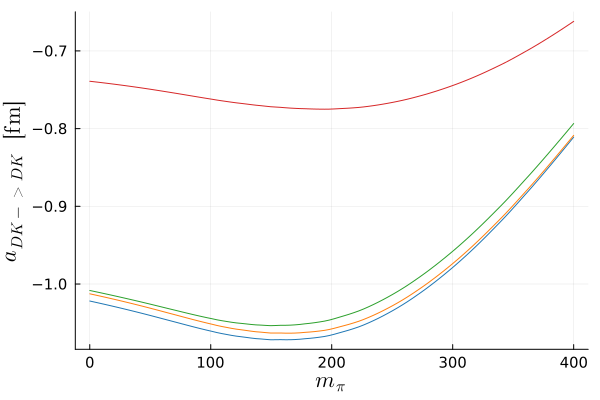}
   \end{minipage}
   \begin{minipage}{0.44\textwidth}
     \centering
     \includegraphics[width=1.0\linewidth]{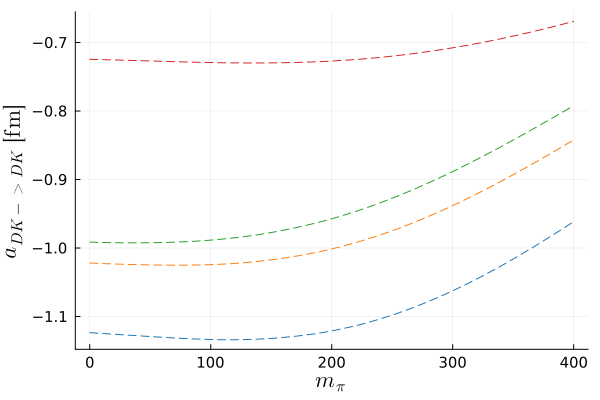}
   \end{minipage}
    \begin{minipage}{0.44\textwidth}
     \centering
     \includegraphics[width=1.0\linewidth]{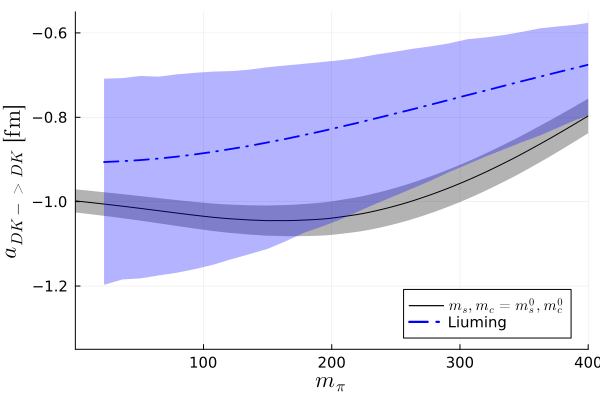}
   \end{minipage}
    \begin{minipage}{0.44\textwidth}
     \centering
     \includegraphics[width=1.0\linewidth]{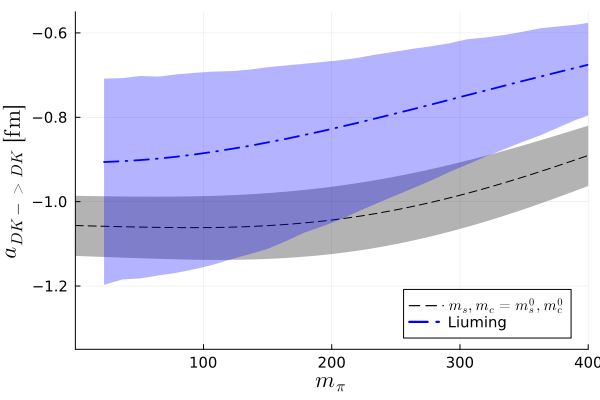}
   \end{minipage}
    \caption{Result of the global fit for the scattering lengths as a function of the pion mass for Fit I (left panels) and Fit II (right panels), in comparison with the result obtained in~\cite{liuorginos} (dashed line with blue band).}
    \label{fig:scattlen1}
\end{figure*}

In Fig.~\ref{fig:1mz1} and \ref{fig:scattlen1} we also show the pion mass dependence of the compositeness and the scattering lengths, respectively, for both fits. The compositeness is a bit lower for the Prelovsek data set, $1-Z\sim 0.6-0.7$, while it gets a value $0.7-0.8$ for all the other collaborations. The scattering lengths are in reasonably good agreement with previous works~\cite{alberto,liuorginos}. However, the errors extracted here are smaller. The scattering lengths obtained here are lower than in \cite{liuorginos}. The reason behind the improvement of the uncertainty is that, on one side, the LQCD data analyzed here are much more precise than the ones for scattering lengths of \cite{liuorginos}. In particular, the data of \cite{cheungthomas} present a great improvement in precision and also the covariance matrix of the energy levels has been taken into account. On the other side, we have included a larger amount of data than in \cite{liuorginos}, where there are $20$ points, while we have $54$ data for the global fit.


\begin{figure}[h!]
    \centering
    \begin{minipage}{0.461\textwidth}
     \centering
     \includegraphics[width=1.0\linewidth]{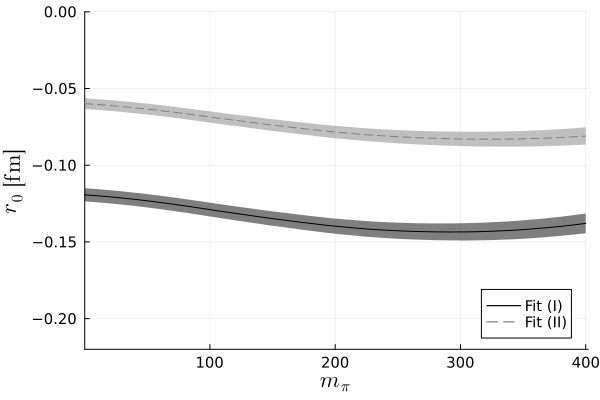}
   \end{minipage}
    \caption{Result of the global fits for the effective range $r_0$ as a function of the pion mass for Fit I and II.}
    \label{fig:r0}
\end{figure}

The effective range is depicted in Fig.~\ref{fig:r0} for both fits. We obtain that, while the scattering length in Fits I and II are compatible, the effective range is more negative for Fit I. These observables are sensitive to the type of potential used. 

\begin{figure}[h!]
    \centering
    \begin{minipage}{0.461\textwidth}
     \centering
     \includegraphics[width=1.0\linewidth]{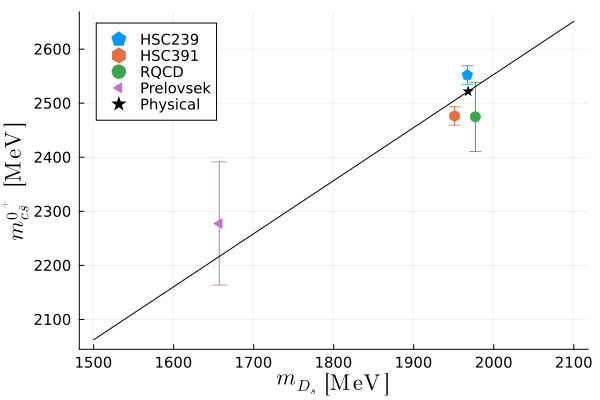}
   \end{minipage}
    \caption{Values of the $m_{c\bar{s}}$ CQM bare masses from our fit as a function of the $D_{s}$ meson mass.}
    \label{fig:mcsline}
\end{figure}

 In Fig.~\ref{fig:mcsline} we show the value of bare mass as a function of the $D_s$ mass for every ensemble. The overall result is compatible with a constant mass difference: $m_{D_s}-m_{D_s^0}=m_{c\bar{s}}-m_{c\bar{s}^0}$, where $m_{c\bar{s}^0}$ and $m_{D_s^0}$ are the values at the physical point. This is in agreement with the assumption done in \cite{pedro}.
 
\begin{figure}[h!]
    \centering
   \begin{minipage}{0.44\textwidth}
     \centering
     \includegraphics[width=1.0\linewidth]{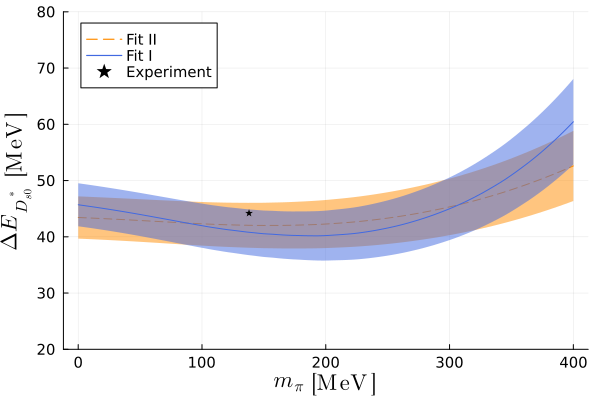}
   \end{minipage}
   \centering
   \begin{minipage}{0.44\textwidth}
     \centering
     \includegraphics[width=1.0\linewidth]{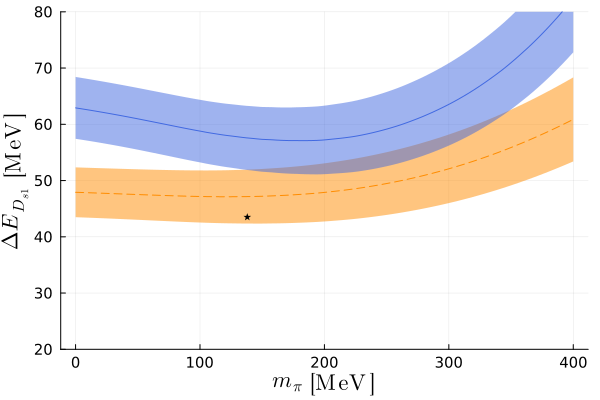}
   \end{minipage}
   \begin{minipage}{0.44\textwidth}
     \centering
     \includegraphics[width=1.0\linewidth]{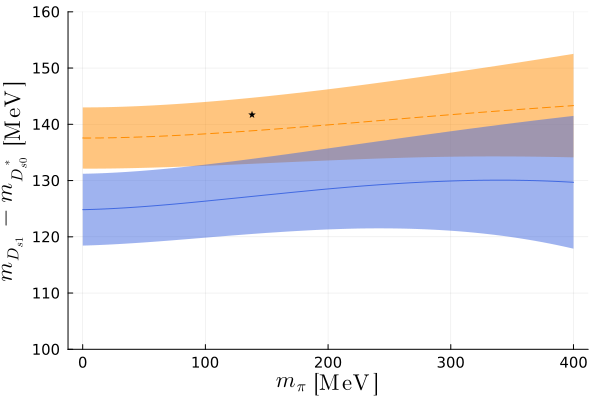}
   \end{minipage}
    \caption{Binding energy of the $D_{s0}^*(2317)$ (top), $D_{s1}(2460)$ (middle) and $1S$ hyperfine splitting (bottom) for physical charm and strange quark masses.}
    \label{fig:final}
\end{figure}

Finally, in Fig.~\ref{fig:final} we plot the binding energies and mass splitting of the $D_{s0}^*(2317)$ and $D_{s1}(2460)$ states as a function of the pion mass in both fits for a physical charm quark mass. To plot Fig.~\ref{fig:final} we have taken the extrapolation of the bare mass (black star) to the physical point from Fig.~\ref{fig:mcsline}.
The binding energies of the $D_{s0}^*(2317)$ and $D_{s1}(2460)$ are compatible with the experimental data for Fit II, while for the Fit I we predict the $D_{s1}(2460)$ state with around $15$ MeV difference from the physical value. They also increase with the pion mass for pion masses above $200$ MeV, showing a quadratic behaviour. The 1S hyperfine splitting is practically constant with the pion mass. 

Finally, in the next section, we evaluate the isospin breaking decay rate $\Gamma(D_{s0}^*(2317)\to D_s^+\pi^0)$.

\newpage

\subsection{Decay width of the $D_{s0}^*(2317)$ to $D_s^+\pi^0$}
\begin{figure}[h!]
   \begin{minipage}{0.23\textwidth}
     \centering
     \includegraphics[width=1.1\linewidth]{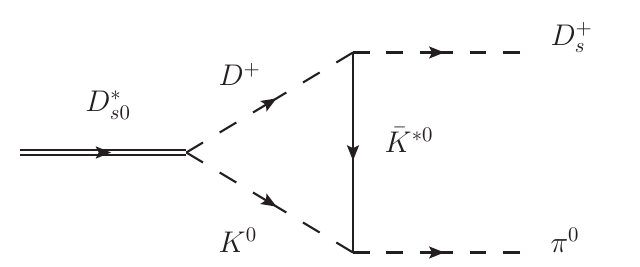}
   \end{minipage}
   \begin{minipage}{0.23\textwidth}
     \centering
     \includegraphics[width=1.1\linewidth]{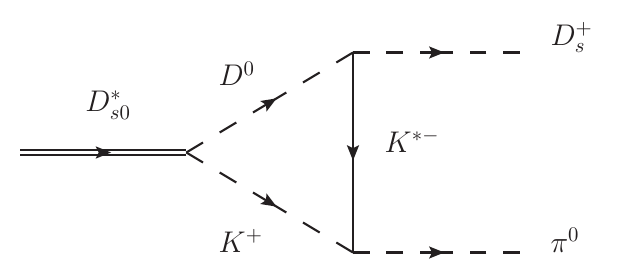}
   \end{minipage} 
   \begin{minipage}{0.23\textwidth}
     \centering
     \includegraphics[width=1.1\linewidth]{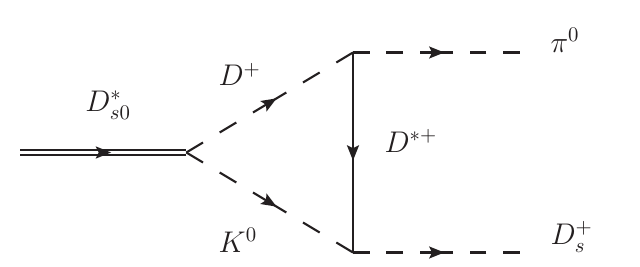}
   \end{minipage} 
   \begin{minipage}{0.23\textwidth}
     \centering
     \includegraphics[width=1.1\linewidth]{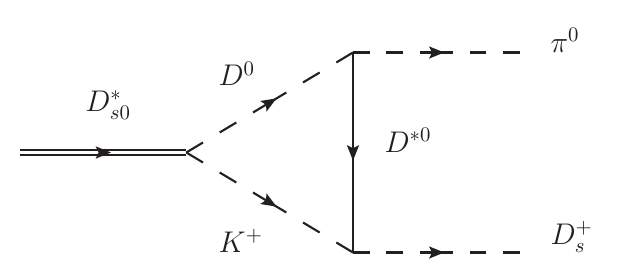}
   \end{minipage} 
   \caption{Feynman diagrams of the $D_{s}^{*0}\rightarrow D_s^+\pi^0$ decay.}
   \label{fig:Ds0toDspi}
\end{figure}

\begin{figure}[h!]
   \begin{minipage}{0.4\textwidth}
     \centering
     \includegraphics[width=1.\linewidth]{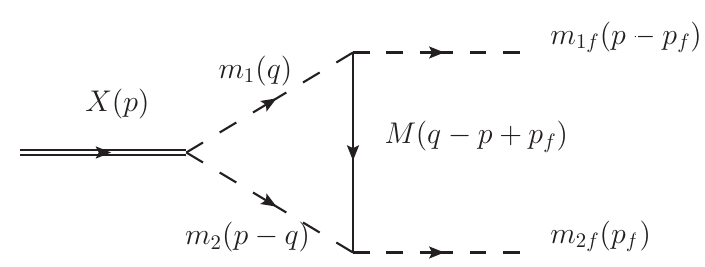}
   \end{minipage}
   \caption{General Feynman diagram of the $D_{s0}^{*}\rightarrow D_s^+\pi^0$ decay.}
   \label{fig:Generaldiagram}
\end{figure}

In this section we evaluate the decay width of the isospin violating decay mode $D_{s0}^*(2317)\rightarrow D_s^+\pi^0$. This process can occur via direct decay as depicted in Fig.~\ref{fig:Ds0toDspi}, 
where one has $\bar{K}^*$ or $D^{*}$ exchange. Besides that, this decay can go through $\pi^0-\eta$ mixing. The diagrams are the same than in Fig.~\ref{fig:Ds0toDspi} but instead of $\pi^0$ we have $\eta_8$ which mixes with the pion through the mixing angle, that appears in the definition of mass eigenstates, 
\begin{eqnarray}
    &\tilde{\pi}^0 = \pi^{0} \cos\tilde{\epsilon} + \eta_8 \sin\tilde{\epsilon} \\
    &\tilde{\eta} = -\pi^{0} \sin\tilde{\epsilon} + \eta_8 \cos\tilde{\epsilon} \nonumber
\end{eqnarray}
with $\eta_8=\frac{2\sqrt{2}}{3}\eta-\frac{1}{3}\eta'$, and is given by,
\begin{eqnarray}
    \tilde{\epsilon}=\frac{\sqrt{3}}{4}\frac{m_d-m_u}{m_s-\hat{m}}\ .
\end{eqnarray}
In the above equation, $\hat{m}$ is the average mass of the $u$ and $d$ quarks. 
We proceed to evaluate the diagram of Fig.~\ref{fig:Generaldiagram} where $X$ is the $D_{s0}^*(2317)$, and $m_1, m_2$ and $m_3$ are the masses of the three pseudoscalars.
The vertices are evaluated by means of the Lagrangian of Eq.~(\ref{eq:lag}). The amplitude of the diagram of Fig.~\ref{fig:Generaldiagram}, is given by,
\begin{eqnarray}\label{eq:intgen}
    &&t=\int \frac{d^4q}{(2\pi)^4} \\
    &&  \frac{g_X g^2 c_1 (p-p_f+q)(p+p_f-q)}{\left[(p-q)^2-m_2+i\epsilon\right]\left[(q-p+p_f)^2-M+i\epsilon\right](q^2-m_1+i\epsilon)} \nonumber
\end{eqnarray}
where $g_X$ is the coupling of the $D_{s0}^*(2317)$ to $DK$, which can be obtained straightforward. Since,
\begin{eqnarray}
     \left|DK, I=0\right> &=& \frac{1}{\sqrt{2}} D^+K^0 + \frac{1}{\sqrt{2}} D^0K^+, 
\end{eqnarray}
one has, $g_X=\frac{g_{DK}^{(I=0)}}{\sqrt{2}}$.
\begin{table}[h!]
 \setlength{\tabcolsep}{0.8em}
{\renewcommand{\arraystretch}{1.6}
\centering
\begin{tabular}{|c|c|c|c|c|c|}
\hline
 $c_1$ & $m_1$ & $m_2$ & $M$ & $m_{1f}$ & $m_{2f}$ \\
 \hline
$-1/\sqrt{2}$ & ${D^+}$ & $K^0$ & ${\bar{K}^{0*}}$ & $D_s^+$ & ${\pi^0}$ \\
 \hline
 $1/\sqrt{2}$ & ${D^0}$ & ${K^+}$ & ${K^{-*}}$ & ${D_s^+}$ & ${\pi^0}$ \\
 \hline
 $-1/\sqrt{2}$ & ${D^+}$ & ${K^0}$ & ${\bar{D}^{+*}}$ & ${\pi^0}$ & ${D_s^+}$ \\
 \hline
  $1/\sqrt{2}$ & ${D^0}$ & ${K^+}$ & ${D^{0*}}$ & ${\pi^0}$ & ${D_s^+}$ \\
 \hline
$\tilde{\epsilon}\sqrt{\frac{3}{2}}$ & ${D^+}$ & ${K^0}$ & ${\bar{K}^{0*}}$ & ${D_s^+}$ & ${\eta}$ \\
 \hline
 $\tilde{\epsilon}\sqrt{\frac{3}{2}}$ & ${D^0}$ & ${K^+}$ & ${K^{-*}}$ & ${D_s^+}$ & ${\eta}$ \\
 \hline
 $\tilde{\epsilon}\frac{1}{\sqrt{6}}$ & ${D^+}$ & ${K^0}$ & ${\bar{D}^{+*}}$ & ${\eta}$ & ${D_s^+}$ \\
 \hline
  $\tilde{\epsilon}\frac{1}{\sqrt{6}}$ & ${D^0}$ & ${K^+}$ & ${D^{0*}}$ & ${\eta}$ & ${D_s^+}$ \\
 \hline
\end{tabular}}
\caption{Masses and coefficients for each diagram that contributes to the $D_{s0}^{*}\rightarrow D_s^+\pi^0$ decay.}
\label{tab:c1masses}
\end{table}

We assume that both couplings, $g_{X,D^+K^0}=g_{X,D^0K^+}=g_X$ are equal, and take $g_{DK}$ from the analysis of LQCD data, where isospin violation is not considered.
We have also approximated $\sum \epsilon_\mu \epsilon_{\nu} \simeq -g_{\mu\nu}$, the constant $c_1$ in Eq.~(\ref{eq:intgen}) is given in Table \ref{tab:c1masses} for every diagram in Fig.~\ref{fig:Ds0toDspi}.
We integrate in the $q^0$ variable applying the Residue Theorem. To make it easier, we decompose the propagators in the positive and negative energy parts. For the two diagramas in the top of Fig.~\ref{fig:Ds0toDspi}, the propagator of $D$ meson ($m_1=m_D$) is,
\begin{eqnarray}
    \frac{1}{q_0^2-\vec{q}\,^2-m_1+i\epsilon}&=&\frac{1}{2\omega_1}\left( \frac{1}{q_0-\omega_1+i\epsilon}-\frac{1}{q_0+\omega_1-i\epsilon}\right)\ ,\nonumber\\ 
\end{eqnarray}
where $\omega_i=\sqrt{\vec{q}\,^2_i+m_i^2}$. Since this is a heavy particle, we can safely neglect the negative energy part of the propagator. Finally, we obtain,
\begin{widetext}
\begin{eqnarray}
    t=&& \frac{g_X g^2c_1}{32\pi^2} \int_0^{q_{max}}\frac{ d^3q}{\omega_1\omega_2\omega_{M}}\left\{\frac{m_X^2-(p_f^0-M_X+\omega_2)^2+(\vec{p}_f-\vec{q})^2}{(M_X-\omega_1-\omega_2+i\eps)(p_f^0-\omega_1-\omega_2)}+\frac{1}{M_X-p_f^0-\omega_1-\omega_M+i\eps)}\right. \\&&\times\left.\left(\frac{M^2_X-(p_f^0-\omega_1)^2+(\vec{p}_f-\vec{q})^2}{M_X-\omega_1-\omega_2+i\eps}-\frac{M_X^2-(2p_f-M_X+\omega_M)^2+(\vec{p}_f-\vec{q})^2}{p_f^0+\omega_M+\omega_2}\right)\right\} \nonumber
\end{eqnarray}
\end{widetext}
being $p_f$ the four-momentum of the pion, and $\omega_M=\sqrt{\vec{q}\,^2+M^2}$. In the integral we have taken $q_{max}=795$ MeV, which corresponds to the same value for $Re(G^{DR})$ in Eq.~\ref{eq:dim} at threshold for $\alpha=-1.91$ MeV, the result of the global fit. For the diagrams in the bottom of Fig.~\ref{fig:Ds0toDspi}, one makes the substitution in the above formula of  $p_{f}^0\rightarrow m_X-p_{f}^0$ and $\vec{p}_{f}\rightarrow-\vec{p}_{f}$.

Finally, we can estimate the width of the $D_{s0}^*(2317)$ using the formula
\begin{eqnarray}
    \Gamma_{X} = |\vec{p}_f|\frac{|t|^2}{8\pi m_{X}^2}.
\end{eqnarray}
We obtain $\Gamma_{D_{s0}^*}=128 \pm 40$ KeV, which supports the molecular interpretation, being in agreement with previous calculations~\cite{liuorginos,Fudecay,Clevendecay}. The error has been estimated varying the couplings in the vertices involving $D^*$ mesons of the diagrams in Fig.~\ref{fig:Ds0toDspi}, from $g=4.20$ to $g=8.95$.

\section{Conclusions}
We have analyzed the available LQCD data on $D^{(*)}K$ scattering \cite{lang,mohler,bali,cheungthomas}. In particular, the HSC energy levels with different boost extracted recently~\cite{cheungthomas}. The formalism is based on the local hidden gauge lagrangian, which has been used in the past to predict several exotic states~\cite{molinabranz,Wu1,Wu2} observed in the past years~\cite{penta1,penta2,LHCb1,LHCb2,tcsbar2900}. The LO term from the HGF is supplemented by a term that accounts for the coupling of the $DK$ channel to a bare $c\bar{s}$ component, respecting both, chiral and heavy quark spin symmetries. 

We have compared the results of the analysis of HSC data with previous LQCD analysis, and with experiment. The results of the HSC data analysis done here agree reasonably well with experiment, and with other previous LQCD studies on $DK$, suggesting the possibility of a global fit. We have conducted this fit and extracted for the first time, both, the quark mass dependence -light and heavy- of the $D_{s0}^*(2317)$ resonance properties from the available LQCD data. In this fit we also include LQCD data for $D^*K$ scattering from other simulations. Our results suggest that the attractive $D^{(*)}K$ interaction increases with the pion mass from $m_\pi\simeq 200$~MeV, but reduces with the charm quark mass. Since most of the discretization error comes from the charm quark, more data on $DK$ scattering at lower than the physical charm quark mass would be needed to confirm this prediction. We obtain that the $D_{s0}^*(2317)$ and $D_{s1}(2460)$ are predominantly molecular. Indeed, the global fit of LQCD data suggest that the properties of these states are sensitive to the light pion mass, reinforcing this statement. However, in comparison with~\cite{Clevenmass}, the dependence obtained for the masses of these resonances is only quadratic for very low pion masses, while it becomes linear when the pion mass increases. We have also improved the precision of the $DK$ scattering lengths extracted previously~\cite{lang,mohler,bali,pedro,alberto,liuorginos}, and studied its quark mass dependence. Compared to~\cite{liuorginos} we obtain smaller values for the scattering lengths. Finally, we obtained a decay width of $\Gamma_{D_{s0}^*\to D_s^+\pi^0}=128 \pm 40$ KeV, in agreement with previous works~\cite{liuorginos,Fudecay,Clevendecay}, and also supporting the molecular interpretation. Future LQCD data analyses for different pion masses than the ones included in our fit, could be a good test for the study done here. 

\section{Acknowledgments}

R. Molina acknowledges support from the CIDEGENT program with Ref. CIDEGENT/2019/015,
the Spanish Ministerio de Economia y Competitividad and European Union (NextGenerationEU/PRTR) by the grant with Ref. CNS2022-13614, and from the spanish national grant PID2020-112777GB-I00. This project has received funding from the European Union’s Horizon 2020 programme No. 824093 for the STRONG-2020 project.

\newpage

\section{Appendix}
\subsection{Matrices $\phi$ and $V$ in Eq.~(\ref{eq:lag})}
\begin{eqnarray}
 &&\phi=\left(
\arraycolsep=0.1pt\def\arraystretch{1.8}\begin{array}{cccc}\frac{\pi^0}{\sqrt{2}}+\frac{\eta}{\sqrt{3}}+\frac{\eta'}{\sqrt{6}}&\pi^+&K^+&\bar{D}^0\\
             \pi^-&-\frac{\pi^0}{\sqrt{2}}+\frac{\eta}{\sqrt{3}}+\frac{\eta'}{\sqrt{6}}&K^0&D^-\\
             K^-&\bar{K}^0&-\frac{\eta}{\sqrt{3}}+\sqrt{\frac{2}{3}}\eta'&D^-_s\\
             D^0&D^+&D_s^+&\eta_c
\end{array}\right)\ ,\nonumber\\
&&V_\mu=\left(\arraycolsep=0.5pt\def\arraystretch{1.8}\begin{array}{cccc}
\frac{\rho^0}{\sqrt{2}}+\frac{\omega}{\sqrt{2}}&\rho^+&K^{*+}&\bar{D}^{*0}\\\rho^-&-\frac{\rho^0}{\sqrt{2}}+\frac{\omega}{\sqrt{2}}&K^{*0}&D^{*-}\\
K^{*-}&\bar{K}^{*0}&\phi&D^{*-}_s\\D^{0*}&D^{*+}&D^{*+}_s&J/\psi\end{array}\right)_\mu\ ,\nonumber\\
\end{eqnarray}
where we have considered ideal $\eta-\eta'$ mixing \cite{Bramon}.

\newpage
\subsection{Figures of HSC energy levels for different boosts}\label{sec:boosts}

\begin{figure}[h!]
    \centering
   \begin{minipage}{0.44\textwidth}
     \centering
     \includegraphics[width=1.0\linewidth]{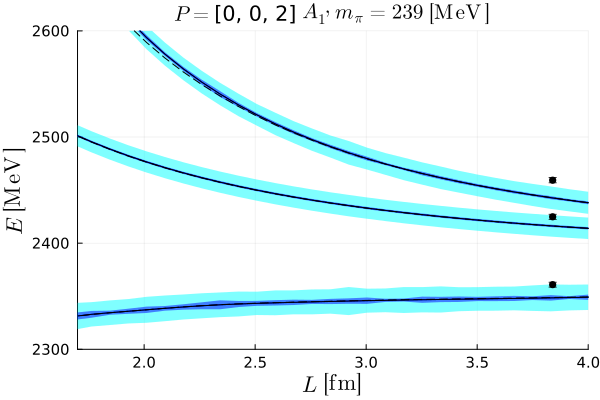}
   \end{minipage}
   \begin{minipage}{0.44\textwidth}
     \centering
     \includegraphics[width=1.0\linewidth]{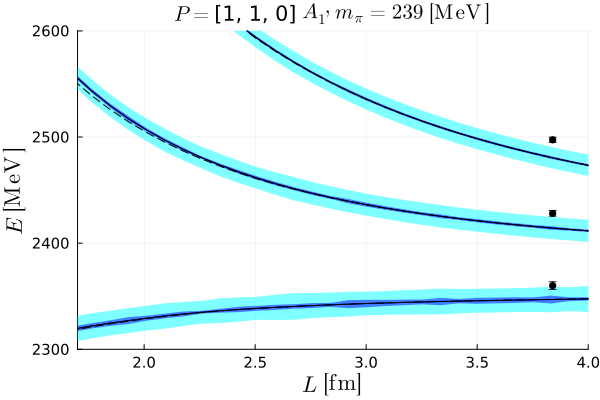}
   \end{minipage}
   \begin{minipage}{0.44\textwidth}
     \centering
     \includegraphics[width=1.0\linewidth]{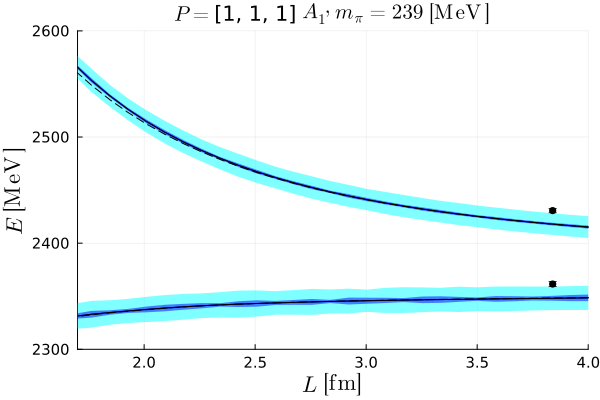}
   \end{minipage}
    \caption{Result from the fit for the dependence with the spatial extent (L) of HSC energy levels for $m_\pi=239$ MeV.}
    \label{fig:energylvlshsc}
\end{figure}

\begin{figure}[h!]
    \centering
   \begin{minipage}{0.44\textwidth}
     \centering
     \includegraphics[width=1.0\linewidth]{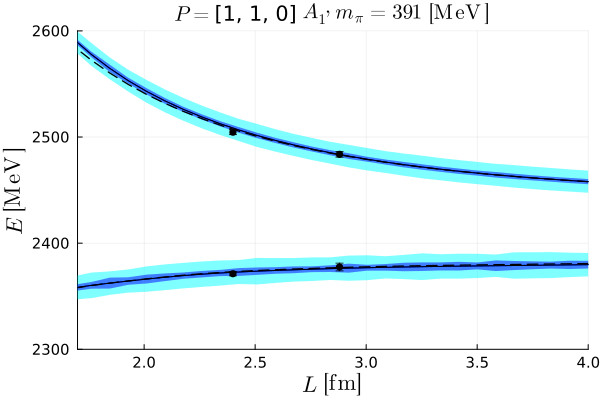}
   \end{minipage}
   \begin{minipage}{0.44\textwidth}
     \centering
     \includegraphics[width=1.0\linewidth]{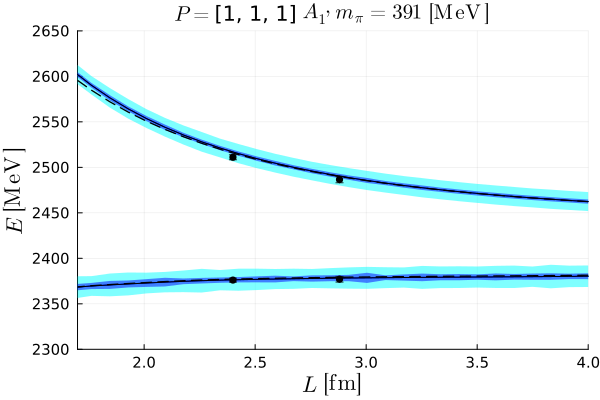}
   \end{minipage}
    \caption{Result from the fit for the dependence with the spatial extent (L) of HSC energy levels for $m_\pi=391$ MeV.}
    \label{fig:energylvlshsc1}
\end{figure}

\newpage
\bibliography{biblio}

\end{document}